 \documentclass[preprint]{aastex}

\slugcomment{Submitted to AJ, 31 May 2000.}

\shorttitle{Thilker, Braun, and Walterbos}
\shortauthors{HIIphot}

\begin{document}


\title{{\bf{\rm H{\small II}{\it phot}}}: AUTOMATED PHOTOMETRY OF \ion{H}{2} REGIONS \\ APPLIED TO M51}


\author{David A. Thilker\altaffilmark{1,2}}
\affil{Department of Astronomy, New Mexico State University, Las
Cruces, NM 88003}
\affil{National Radio Astronomy Observatory, Socorro, NM 87801}
\email{dthilker@nrao.edu}

\author{Robert Braun}
\affil{Netherlands Foundation for Research in Astronomy, \\
P.O. Box 2, 7990 AA Dwingeloo, The Netherlands}
\email{rbraun@nfra.nl}

\and

\author{Ren\'e A. M. Walterbos\altaffilmark{3}}
\affil{Department of Astronomy, New Mexico State University, Las
Cruces, NM 88003}
\email{rwalterb@nmsu.edu}


\altaffiltext{1}{present address: National Radio Astronomy
Observatory, P.O. Box O, 1003 Lopezville Road, Socorro, NM 87801}
\altaffiltext{2}{Jansky Fellow}
\altaffiltext{3}{Visiting Astronomer, Kitt Peak National Observatory.
KPNO is operated by AURA, Inc.\ under contract to the National Science
Foundation.}


\begin{abstract}
We have developed a robust, automated method, hereafter designated
{\bf{\rm H{\small II}{\it phot}}}, which enables accurate photometric
characterization of \ion{H}{2} regions while permitting genuine adaptivity
to irregular source morphology.  {\bf{\rm H{\small II}{\it phot}}}
utilizes object-recognition techniques to make a first guess at the
shapes of all sources, then allows for departure from such idealized
``seeds'' through an iterative growing procedure.  Photometric
corrections for spatially coincident diffuse emission are derived from
a low-order surface fit to the background after exclusion of all
detected sources.  We present results for the well-studied, nearby
spiral M51 in which 1229 \ion{H}{2} regions are detected above the $5\sigma$
level.  A simple, weighted power-law fit to the measured H$\alpha$
luminosity function (\ion{H}{2} LF) above log L$_{H\alpha} = 37.6$ gives
$\alpha = -1.75\pm0.06$, despite a conspicuous break in the \ion{H}{2} LF
observed near L$_{H\alpha} = 10^{38.9}$.  Our best-fit slope is
marginally steeper than measured by Rand (1992), perhaps reflecting
our increased sensitivity at low luminosities and to notably diffuse
objects.  \ion{H}{2} regions located in interarm gaps are preferentially
less luminous than counterparts which constitute M51's grand-design
spiral arms and are best fit with a power-law slope of $\alpha =
-1.96\pm0.15$.  We assign arm/interarm status for \ion{H}{2} regions based
upon the varying surface brightness of diffuse emission as a function
of position throughout the image.  Using our measurement of the
integrated flux contributed by resolved
\ion{H}{2} regions in M51, we estimate the diffuse fraction to be
approximately 0.45 -- in agreement with the determination of Greenawalt
et al. (1998).  Automated processing of degraded narrowband
datasets is undertaken in order to gauge (distance-related) systematic
effects associated with limiting spatial resolution and sensitivity.
\end{abstract}




\section{Introduction}
\label{sec:HIIphotcode.introduction}

\ion{H}{2} regions are an effective optical tracer of ongoing massive star
formation.  Even a single early-type star produces enough Lyman
continuum photons to ionize a quantity of gas sufficient to produce
readily detectable recombination lines in galaxies at distances of
many Mpc.  Observations in the Balmer lines (predominantly
H{$\alpha$}) and other nebular emission lines (e.g. [\ion{N}{2}]
{$\lambda\lambda$} 6548,6584, [\ion{S}{2}] {$\lambda\lambda$} 6717,6731, and
[\ion{O}{3}] {$\lambda$} 5007) enable estimation of physical conditions
within discrete \ion{H}{2} regions (\citet[]{evansdopita85},
\citet[]{osterbrock89} ($AGN^2$), \citet[]{ferlandetal98}).  Conversely, measurement of the \ion{H}{2} region luminosity function
can indicate global patterns in the process of star cluster formation
(\citet[]{kennicuttetal89} (hereafter KEH89),
\citet[]{banfietal93}, \citet[]{feinstein97},
\citet[]{wyderetal97}).  An excellent review of the field is
provided by \citet[]{oeyclarke98}.

We sought a fully-automated technique for determination of the
positions, fluxes, and sizes of \ion{H}{2} regions in galaxies.  Such a
tool is crucial for efficient and reproducible characterization of
their star formation properties, especially if meaningful
intercomparison between datasets is a primary goal.  Another advantage
of an automated approach over conventional measurement by visual
inspection is that many systematic effects, like that of reduced
spatial resolution for more distant galaxies or of differences in
limiting sensitivity, can be quantitatively ascertained.

Well-resolved images of spiral and irregular galaxies reveal that
massive star clusters often form in close proximity, usually leading
to a confused and rather inhomogeneous distribution of ionized gas.
This inherent clumpiness makes automated photometry difficult.
Photometric accuracy can be further hindered by the presence of a
variable background of diffuse ionized gas (DIG).  In complex
environments of this nature it is practically impossible to cleanly
separate the contribution of neighboring extended objects to the
observed surface brightness distribution.  Methods of correcting for
source overlap in the special case of stellar photometry (e.g.
DAOPHOT, \citet[]{stetson87}; ALLFRAME, \citet[]{stetson94}) cannot be
applied without accurate models for the intrinsic structure of every
source.  At present it is infeasible to construct a comprehensive set
of models spanning the observed properties of resolved \ion{H}{2} regions
in nearby galaxies.  {\it Any automated photometric procedure
optimized for \ion{H}{2} regions must consequently provide adaptivity to
the actual source morphology.  This can be accomplished using an
iterative approach to ``grow'' sources from an initial guess at the
shape.}  We have developed {\bf{\rm H{\small II}{\it phot}}}, a
user-friendly procedure written in IDL
\footnote{For information on the Interactive Data Language (IDL), see
  http://www.rsinc.com.} which employs such a method.

\citet[]{mccalletal96} demonstrated the potential of an
automated photometry procedure for \ion{H}{2} regions based on a simple
iterative growth mechanism.  Their method, called percent-of-peak
photometry (PPP), involved growth from local maxima down to a constant
fraction of the difference between each peak and its local background.
In the grand-design spiral NGC~3398 and the flocculent spiral
NGC~4414, PPP successfully reproduced the LF obtained through standard
fixed-threshold photometry (FTP).  \citet[]{kingsburghmccall98} have
recently applied PPP during their analysis of four nearby dwarf
galaxies.  Unfortunately, McCall and collaborators are unable to
recover more than a small percentage of the observed flux for even the
brightest sources when using PPP.  This stems from the fact that they
can only grow to 70\% of peak before the automated method becomes
susceptible to rapid growth and merging of adjacent regions.  Our
present research was inspired by the desire to overcome this
limitation using criteria to carefully regulate growth in saddle
points between neighboring regions.  Also, we hoped to recover all the
observed flux rather than only that contributed by each region's
brightest pixels by defining larger ``seeds'' with a better match in
shape to the source structure (rather than growing from a single
pixel).  McCall et al. argued that the flux detected using PPP was
directly proportional the total source flux, but their line of
reasoning assumed an idealized Stromgren sphere geometry for all
regions.  It is difficult to imagine that this assumption could be
satisfied in general.

This paper is organized in the following manner.
Section~\ref{sec:HIIphotcode.procedure} presents the concepts and
algorithms employed within {\bf{\rm H{\small II}{\it phot}}}.
Section~\ref{sec:HIIphotcode.M51.obs} contains a very brief
description of the M51 dataset used for illustrating the capabilities
of the procedure.  Section~\ref{sec:HIIphotcode.results} presents the
population of \ion{H}{2} regions in M51, including a new, more sensitive
luminosity function (LF). Finally, we conclude in
Section~\ref{sec:HIIphotcode.summary} with a summary and view toward
the near future.

\section{{\bf{\rm H{\small II}{\it phot}}} procedure}
\label{sec:HIIphotcode.procedure}

{\bf{\rm H{\small II}{\it phot}}} is a completely automated method for
photometry of \ion{H}{2} regions.  Our algorithm is sufficiently general to
work well for distant galaxies, but provides the most substantial
benefit during analysis of narrowband images of complicated,
highly-resolved systems.  Below is an explicit description of the
procedure.

\subsection{Initial detection of sources}
\label{sec:HIIphotcode.initial.detection}

Following the discussion of astronomical object recognition recently
presented by \citet[]{thilkeretal98},
hereafter TBW98, \citet[]{thilker98}; and
\citet[]{mashchenkoetal99}, hereafter
MTB99, we recognize that an ideal technique for decomposition of
narrowband images into individual objects might employ: (1)
calculation of projected physical models describing all anticipated
source morphologies, (2) cross-correlation of image data with each
model to find tentative matches, and finally (3) pruning of the
composite detection list to correct for multiple detections of the
same source.  Regrettably, this direct approach is not currently
viable due to a lack of sufficient computing power and a comprehensive
set of models.  One practical alternative might be to select a set of
sufficiently diverse empirical models and evaluate the degree to which
they match the data at some limited set of sky positions.  {\bf{\rm
H{\small II}{\it phot}}} employs this strategy, followed by an
iterative growing procedure to permit departures from the idealized
models.

The {\bf{\rm H{\small II}{\it phot}}} collection of empirical models
includes six basic morphologies, each considered at various sizes and
with major-to-minor axis ratios ranging from one to two, stepping by
0.25.  We permit different position angles, sampling with an increment
of 15\arcdeg.  In each morphology, the predicted surface brightness of
the radially symmetric (base) model is computed as:

\begin{equation}
f(r) = exp~{\frac{-(r-r_0)^2}{2 {\sigma}^2}}.
\end{equation}

\begin{figure}
\centering
\resizebox{\hsize}{!}{\includegraphics{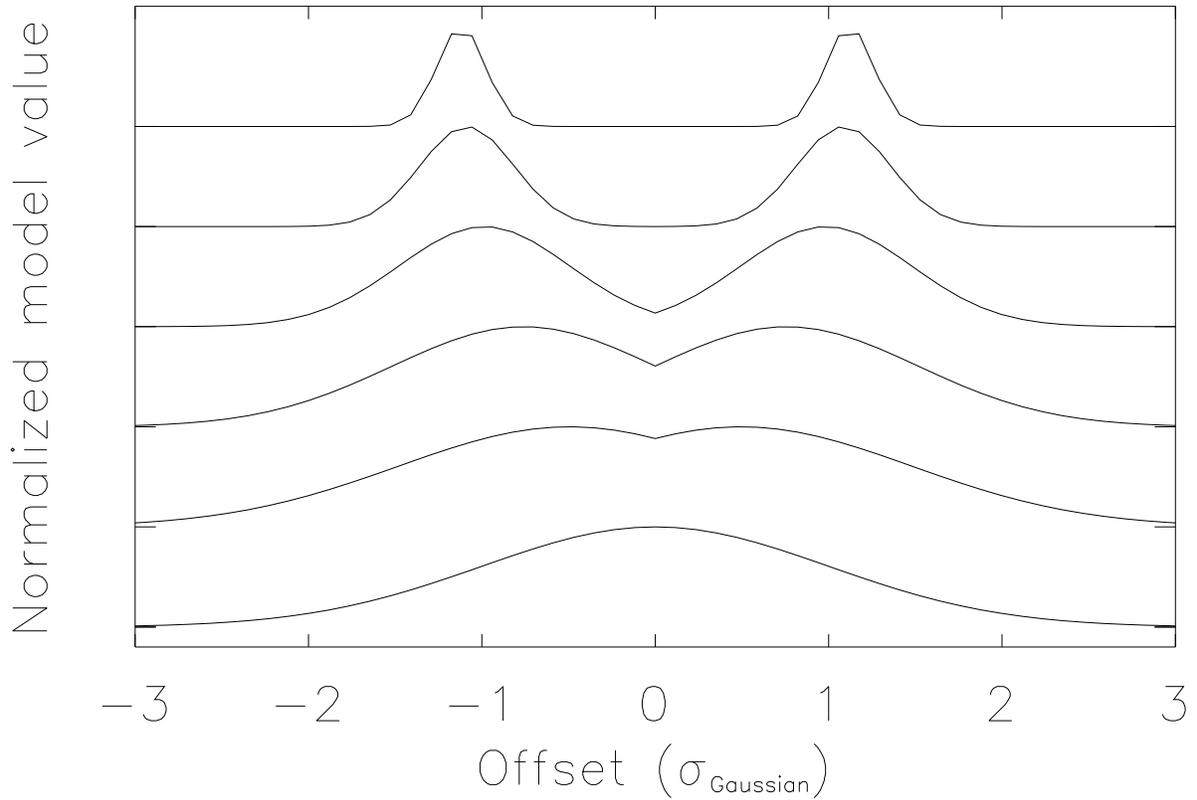}}
\addtocontents{lof}{\setlength{\baselineskip}{\singlespace}
\vspace{\baselineskip}}
\caption[Normalized, model-center cross cuts for each of
  the model types included in our initial search for \ion{H}{2}
  regions]{Normalized, model-center cross cuts for each of the model
  types included in our initial search for \ion{H}{2} regions.  We include
  Gaussians, centrally-depressed structures, and ring morphologies.
  Relative sizes of each model in this figure have been specified so
  as to be associated with a single detection kernel.}
\label{fig:normalized.model.cuts}
\addtocontents{lof}{\setlength{\baselineskip}{\doublespace}}
\end{figure}

Figure~\ref{fig:normalized.model.cuts} shows a model-center cross-cut
for every morphological class.  Each profile has been normalized to
unit peak brightness.  We include Gaussians by setting r$_0$~=~0,
whereas ring models of varied ``shell thickness'' (relative to the
ring diameter) are generated by taking r$_0 > 0$ and adopting various
$\sigma/r_0$ ratios.  Specific choices of $\sigma/r_0$ were selected
in order to sample thin rings, thick rings, and centrally depressed
structures.  Note that in Fig.~\ref{fig:normalized.model.cuts} we
varied r$_0$ with the intent of producing sources having the same
characteristic size.  As mentioned above, each radially symmetric base
model is stretched and rotated in numerous ways for comparison with
the data.

The essential challenge when incorporating these parameterized ``guess
morphologies'' into {\bf{\rm H{\small II}{\it phot}}} is finding a way
to limit the number of sky positions at which any model must be
compared with the data.  Because we use at least 100 stretched/rotated
variants for each base model of a given size, together with typically
50 base model sizes, it is prohibitive to compute a cross-correlation
between each model and the entire image plane (as implemented by TBW98
for the case of \ion{H}{1} shells).  Instead, we determine a list of
``tentative match'' sky coordinates for each model by tabulating
significant local maxima in the convolution of the data with an
appropriately-sized circular Gaussian.  In this manner, we detect
structures with dissimilar morphology but having about the same size
in a single pass.  Our technique works because we only look for
\ion{H}{2} regions of a given characteristic size on images that have been
smoothed to remove source structure on smaller scales.  Even the most
well-resolved ring (for instance) will end up looking like a Gaussian
after some degree of spatial smoothing.  We use ``lowered'' Gaussian
kernels (as also employed in DAOPHOT, \citet[]{stetson87}) as a means of removing
slowly varying background structure from the galaxy image during our
multi-resolution, convolution-based procedure. Each Gaussian kernel was
truncated at a radius of 1.5$\sigma$ and offset with a constant in order
to provide an integral over the kernel of zero.  We tabulate solely
those convolution maxima which have peaks exceeding a $5\sigma$
threshold.  Variance associated with random fluctuations in the
convolution of each Gaussian kernel with our data is measured in a
user-selected sky region.  Modest flat-fielding errors are not problematic
due to our use of a lowered Gaussian as the convolution kernel.

After compiling a list of tentative centroid positions for sources of
each characteristic size, direct comparison between a set of
stretched/rotated models (Gaussians and rings) and the data is
accomplished by calculation of a noise-corrected version of Pearson's
linear correlation coefficient, $\rho$.  (As described in detail by
MTB99, this statistic allows robust estimation of ``goodness of fit''
and completeness.  Note that $\rho$ is invariant under linear scaling
of the data, so the flux of a region and the level of its local
background are irrelevant. Only the best match (highest $\rho$) model
together with it's value of $\rho$ are retained for each tentative
source. The entire list of tentative sources is sorted by the $\rho$
value of the entries.  We then employ a cutoff, $\rho_{crit}$, in the
correlation coefficient in order to retain only the best matches.  For
this paper we adopted $\rho_{crit} = 0.25$, although the median value
was $\sim 0.75$.  Remember that so far we are only creating a ranked
list of possible detections.

\subsection{{\bf{\rm H{\small II}{\it phot}}} footprint and seed definition}
\label{sec:HIIphotcode.foot.seed}

Having this sorted list of potential detections, we next eliminate
multiple detections associated with the same observed emission.  This
is accomplished by defining ``footprints'' in the image for each
source.  Beginning with the highest ranked detection, we loop over all
regions allowing each one to ``claim'' pixels of the input image.
Each detection is allowed to place a footprint if the following
conditions are satisfied: (1) the associated model centroid has not
been claimed, (2) 90\% of the data flux inside the model's 20\%
isophotal boundary remains unclaimed, and (3) the detection's
signal-to-noise is greater than ${\frac{S}{N}}_{crit}$.  (See
Section~\ref{sec:HIIphotcode.flux.and.SN} for a detailed discussion of
signal-to-noise in the context of {\bf{\rm H{\small II}{\it phot}}}.
We introduce a formal analysis based on uncertainties associated with
the independent line+continuum and continuum images, rather than
merely the continuum-subtracted image.)  Regions satisfying these
conditions take as a footprint all unclaimed pixels within the 20\%
isophotal level of their best-match model.  Effectively, our footprint
convention allows simultaneous rejection of multiple detections
(naturally leaving only the best-match model) and introduces a buffer
between neighboring regions.  One can think of this procedure as a
detailed fitting process in which all sources are compared with a
finite number of relatively well-matched models.

Due to line-of-sight projection, one should anticipate overlapping
\ion{H}{2} regions in most galaxies (even if perfectly face-on, due to the
finite disk thickness).  Note that our methodology makes it possible
to separately detect and analyze sources even when they have complete
spatial overlap if their morphologies are sufficiently distinct.  If a
compact, highly significant source first places a footprint in an area
containing many surface brightness enhancements with a range in size,
the probability is substantial that a larger, less significant source
will overlap the initial detection.  This large detection will be
allowed into the catalog provided the compact region does not contain
more than 10\% of the observed flux within the big model's 20\%
isophotal boundary (presuming the other standard conditions for a
footprint are also met).  The {\bf{\rm H{\small II}{\it phot}}}
procedure naturally treats partially overlapping and fully overlapping
detections in this manner.

\begin{figure}
\centering
\resizebox{0.95\hsize}{!}{\includegraphics{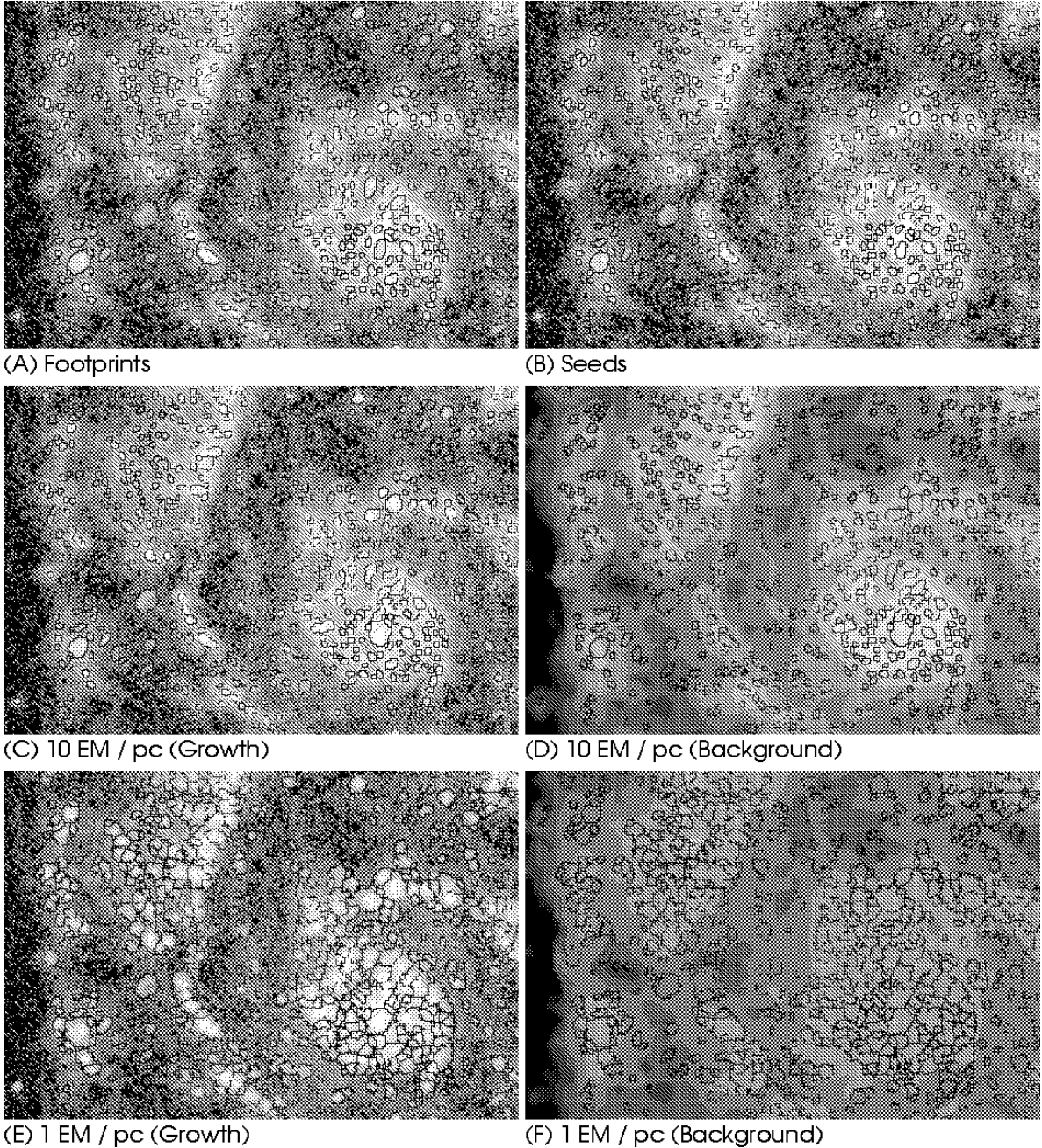}}
\caption[Subsection of M51 shown at various stages of
  the {\bf{\rm H{\small II}{\it phot}}} procedure]{Subsection of M51 shown at various stages of
  the {\bf{\rm H{\small II}{\it phot}}} procedure. Panel (a) contains
  a continuum-subtracted H$\alpha$ image with ``footprint'' boundaries
  indicated.  In panel (b), we present the same image with ``seed''
  boundaries marked.  Panel (c) indicates the extent of each region
  after growth to a terminal surface brightness slope of 10 EM/pc.
  Panel (d) shows a surface fit to the diffuse background emission
  remaining after growth to the state presented in panel (c).  This
  type of fit is used to make corrections to the integrated flux of
  each \ion{H}{2} region.  Panel (e) indicates the maximum extent ever
  allowed by the {\bf{\rm H{\small II}{\it phot}}} growing procedure
  (1 EM/pc terminal slope).  At this point, \ion{H}{2} regions boundaries
  contain not only the classical \ion{H}{2} region but also any associated
  DIG in the immediate vicinity.  Panel (f) is a fit to the
  slowly-varying background emission after growth to the maximum
  extent.}
\label{fig:HIIphot.demo}
\end{figure}

\begin{figure}
\centering
\resizebox{\hsize}{!}{\includegraphics{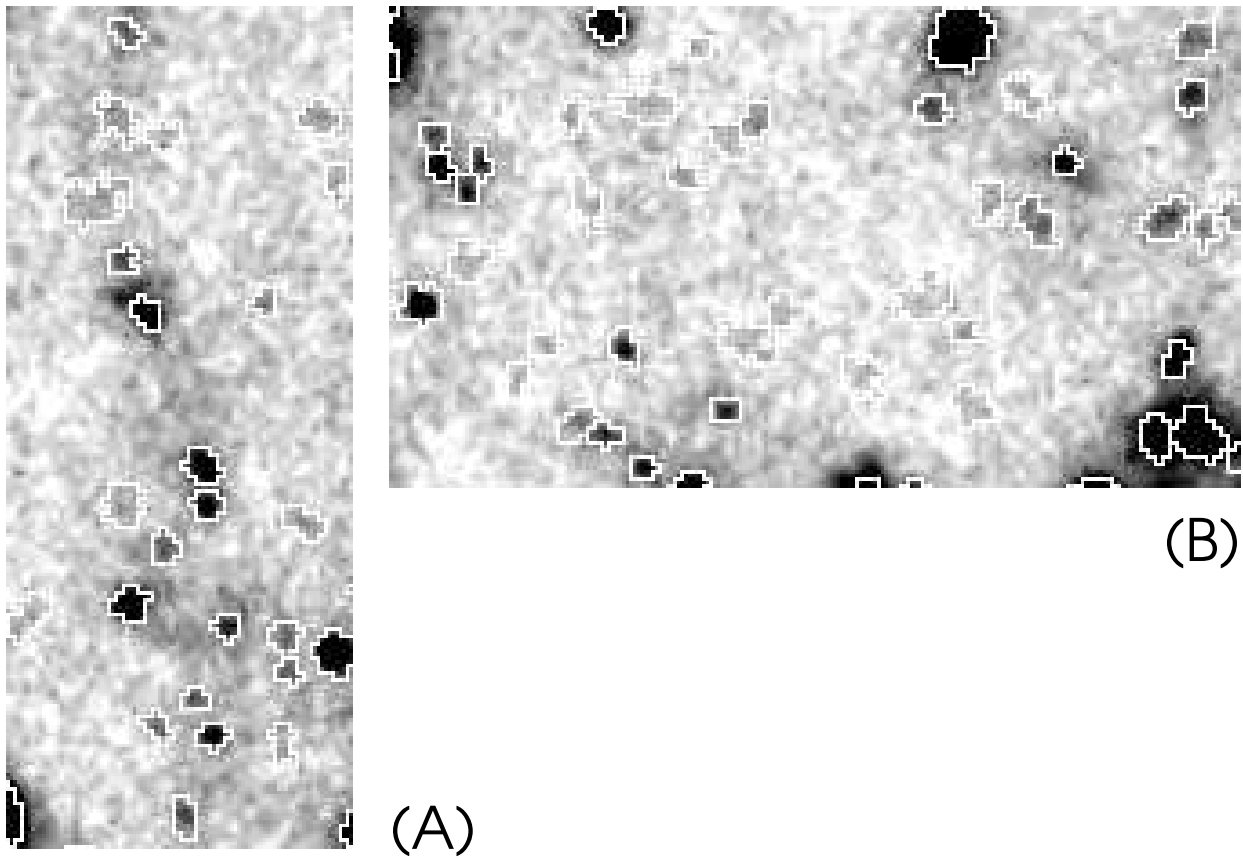}}
\addtocontents{lof}{\setlength{\baselineskip}{\singlespace}
\vspace{\baselineskip}}
\caption[Two small sections of the image presented in
  Fig.~\ref{fig:HIIphot.demo}c, redisplayed with high-contrast linear scaling to better show
  the faintest detected \ion{H}{2} regions]{Two small sections of the image presented in
  Fig.~\ref{fig:HIIphot.demo}c, redisplayed with high-contrast linear scaling to better show
  the faintest detected \ion{H}{2} regions.  As described in the text, this
  figure indicates boundaries for all detections, regardless of their
  {\em final} signal-to-noise ratio.  All such detections do pass a
  $5\sigma$ cut in the multi-resolution FIND routine, but
  signal-to-noise may subsequently drop as regions grow into fainter
  areas for which noise is relatively more significant.  Scientific
  analysis only uses sources for which
  ${\frac{S}{N}}_{final} \ge 5$, unless specifically stated
  otherwise.}
\label{fig:high.contrast}
\addtocontents{lof}{\setlength{\baselineskip}{\doublespace}}
\end{figure}

Figure~\ref{fig:HIIphot.demo} illustrates the complete {\bf{\rm
H{\small II}{\it phot}}} procedure by showing the same image section
within a continuum subtracted H$\alpha$ image of M51 at various stages
of the processing.  In particular, Fig.~\ref{fig:HIIphot.demo}a shows
our {\bf{\rm H{\small II}{\it phot}}} footprints.  The image data has
been scaled logarithmically to keep from saturating the inner portions
of the galaxy.  Scaling is identical in each panel so as to facilitate
comparison between panels
\ref{fig:HIIphot.demo}c--\ref{fig:HIIphot.demo}f.  All marked regions
are associated with a convolution peak ($5 \sigma$ or better) at
either original or somewhat degraded resolution.
Fig.~\ref{fig:high.contrast} shows two small subsections of
Fig.~\ref{fig:HIIphot.demo}c (see description further below) with
linear scaling in order to demonstrate the significance of low surface
brightness detections which are difficult to appreciate in
Fig.~\ref{fig:HIIphot.demo}.  Not all of the detections shown in
Fig.~\ref{fig:high.contrast} are used in the construction of our \ion{H}{2}
region LF, as we demand that every ``photometric source'' have a final
signal-to-noise ratio in excess of five.  Nevertheless, all detections
plotted are thought to be genuine, having been originally discovered
using {\bf{\rm H{\small II}{\it phot}}} and later confirmed by visual
inspection of individual continuum-subtracted images (before
CR-rejection) viewed at various resolutions.  Recall that we never
make use of (and draw no conclusions from) these intrinsically
questionable detections.  Essentially they should be considered
candidates, until deeper observations become available.

Because our empirical models are only a first order approximation to
actual source structure, footprints often contain pixels that are not
bright enough to justifiably remain in the final boundary of the
region.  We account for this by rejecting all pixels which fall
outside a ``bounding isophote'' defined by 50\% of the median data
value found within each footprint (where the median is measured
relative to an estimated local background).  We call these trimmed
footprints ``seeds'' since they are composed of only those pixels
destined to belong to a region, but do not yet reflect changes
associated with the iterative growth procedure.  Our procedure ensures
that all seed boundaries follow isophotal contours within footprint
boundaries, although the specific cutoff varies depending on the
distribution of pixel intensities within any given \ion{H}{2} region.
Notice that this conservative approach makes it possible for ring-like
footprints to reject pixels which fall within the object's central
surface brightness depression.  Fig.~\ref{fig:HIIphot.demo}b shows the
{\bf{\rm H{\small II}{\it phot}}} seeds for our subsection of M51.  In
practice, our particular implementation of the ``seed'' convention is
motivated by the following arguments: (1) multi-pixel seeds provide a
``head start'' for the iterative growth process, making it easier to
reliably separate adjacent regions, (2) defining seeds as a
data-regulated subset of model footprints allows the first true
excursion of region boundaries from our set of empirical models.

\subsection{Iterative growth of detections}
\label{sec:HIIphotcode.iterative.growth}

Given a set of seed pixels associated with every \ion{H}{2} region in a
galaxy, it might seem a simple matter to iteratively add pixels to
each region until reaching the maximum extent of all nebulae. In fact,
the implementation of a well-behaved iterative growing algorithm is
far from trivial and there is no established convention for
determining the ``edge'' of an \ion{H}{2} region.  \citet[]{mccalletal96}
encountered difficulty in growing their sources to isophotal cutoffs
fainter than about 70\% of the local peak.  Potential inhomogeneity in
the diffuse background level and crowding of regions having remarkably
different flux conspire to make the PPP method less suitable except in
a limited set of well-behaved circumstances.  {\bf{\rm H{\small
II}{\it phot}}} attempts to carefully control the rate of growth in
saddle points between regions by introducing a slowly declining
threshold which determines the set of pixels considered for growth
during a given iteration.  Pixels having values below this global
threshold are ignored until later iterations.  In this way,
neighboring regions approach their saddle point at an equal rate no
matter what the difference in peak value or total counts between
sources.

Iterative growth commences by setting the global threshold for pixel
consideration equal to the highest bounding isophote and is reduced by
0.02 dex before each subsequent iteration.  Regions as a whole are
considered for growth only if the median value of the pixels in a
seed's ``exterior perimeter'' exceeds the slowly declining threshold.
This implies that only the seed having the highest bounding isophote
is considered during the first iteration.  Any time that more than one
region is allowed to grow during an iteration, {\bf{\rm H{\small
II}{\it phot}}} cycles through the active regions in order of
decreasing correlation coefficient, $\rho$.  Qualified pixels (lying
above the global threshold) which are adjacent to or diagonal from any
pixel already belonging to the region being augmented are potentially
added to the source if they are not claimed by other regions.  That
is, pixels from the exterior perimeter of a region can be added if
they are bright enough.  We also require that at least 50\% of the
perimeter pixels are added during any given iteration.  If this is not
the case, we postpone growth until the global threshold declines
further, so as to simultaneously add most of an entire isophotal ring.
Growth for a particular region continues in this manner until either:
(1) the observed surface brightness profile flattens sufficiently, or
(2) no more qualified pixels can ever be reached due to being
surrounded by other regions or because of the intrinsic data values.
Note that regions can ``stall'' for many iterations and do not
immediately cease growing just because neighbor pixels cannot be
considered (as a result of the global threshold).  In other words, our
iterative procedure amounts to carefully adding lower isophotal
contours to all qualified regions after specifically accounting for a
slightly unequal start brought about by our adaptive definition of
seed boundaries.

\begin{figure}
\centering
\resizebox{\hsize}{!}{\includegraphics{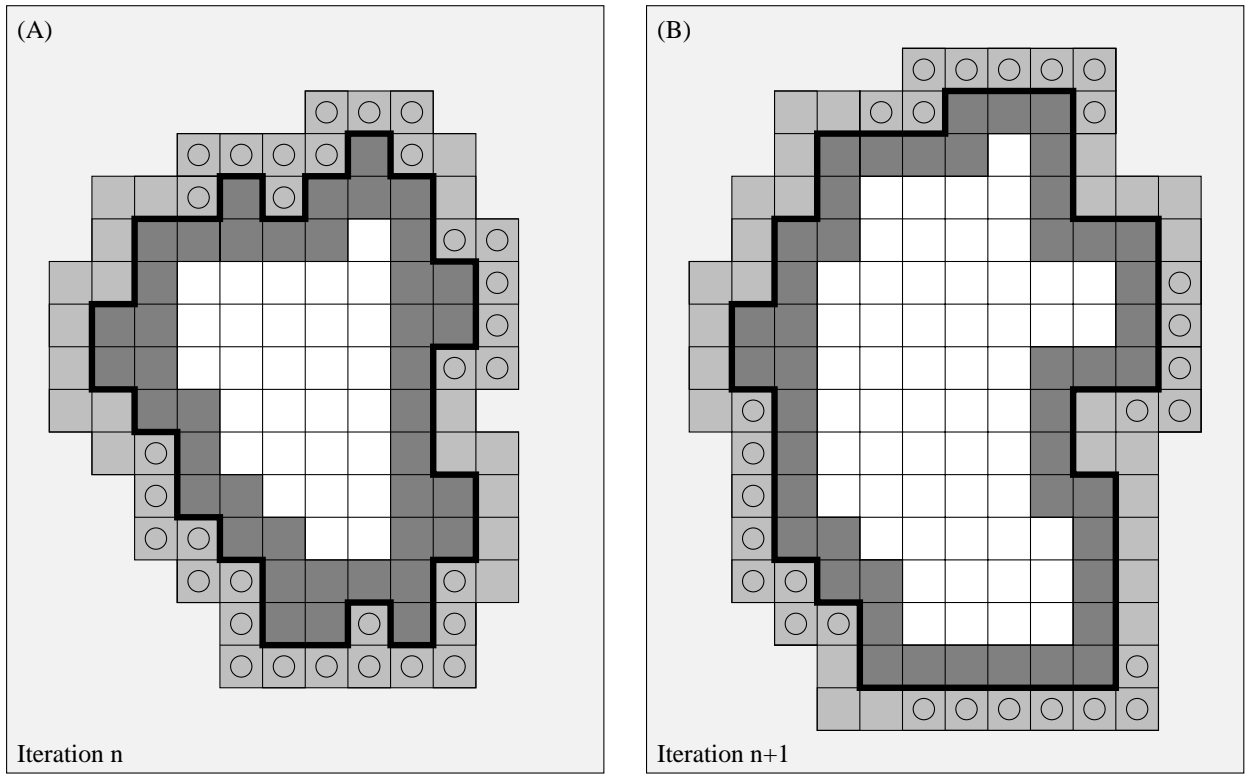}}
\addtocontents{lof}{\setlength{\baselineskip}{\singlespace}
\vspace{\baselineskip}}
\caption[Schematic representation of a hypothetical \ion{H}{2}
  region undergoing iterative growth]{Schematic representation of a hypothetical \ion{H}{2}
  region undergoing iterative growth.  We indicate the instantaneous
  boundary of the source with a heavy solid line.  Pixels belonging to
  the ``interior perimeter'' set are shaded dark, whereas the
  ``exterior perimeter'' set is lightly filled.  Members of the
  exterior set which exceed the global threshold for growth during
  each iteration are marked with circles.}
\label{fig:growth.schematic}
\addtocontents{lof}{\setlength{\baselineskip}{\doublespace}}
\end{figure}

Figure~\ref{fig:growth.schematic} shows schematic representations of a
hypothetical source being considered for iterative growth.  In panel
(4a), the dark shaded pixels belong to the region's interior perimeter
set during active iteration $n$, whereas lighter colored pixels
compose the exterior perimeter group.  The current region boundary is
indicated with a heavy solid line.  Median values for the
interior/exterior perimeter sets will be used to determine if the
surface brightness profile has flattened sufficiently in order to stop
further growth (in subsequent iterations).  Some of the lightly shaded
pixels have been marked with a circle.  These exterior perimeter
pixels have values above the global threshold and will be added to the
region during iteration $n$.  Note that more than 50\% of the lightly
shaded pixels fall into this category.  If this had not been the case,
growth for this region would stall until the {\bf{\rm H{\small II}{\it
phot}}} global threshold declined enough to allow a majority of the
exterior perimeter pixels to augment the region.  Panel
(\ref{fig:growth.schematic}b) is similar to Panel
(\ref{fig:growth.schematic}a), except that it has been drawn for the
following iteration, $n+1$.

The question of how to determine whether a surface brightness profile
has ``flattened'' is somewhat difficult to treat on anything other
than pragmatic grounds.  Presently there is no established connection
between the rate of surface brightness decline and specific physical
conditions within an \ion{H}{2} nebula.  Originally we demanded that
regions grow until the difference in median values between interior
and exterior perimeter pixel sets indicated the surface brightness
profile was no longer declining.  This choice resulted in very large
\ion{H}{2} regions and significant bumping of adjacent regions, since in
crowded fields brightness profiles rarely flatten out before
encountering a neighbor.  Our reason for requiring that the surface
brightness profile flatten completely was that we sought to fairly
treat all regions, regardless of their environment.  Using this
procedure we effectively determined groups of pixels most plausibly
associated with the same ionizing source.  That is, for this
methodology, our \ion{H}{2} ``regions'' included compact cores and related
diffuse emission (DIG).  Although interesting in its own right, this
non-conventional definition of an \ion{H}{2} region makes it difficult
to compare the current results with previous work and we sought a more
flexible alternative.

\begin{figure}
\centering
\resizebox{\hsize}{!}{\includegraphics{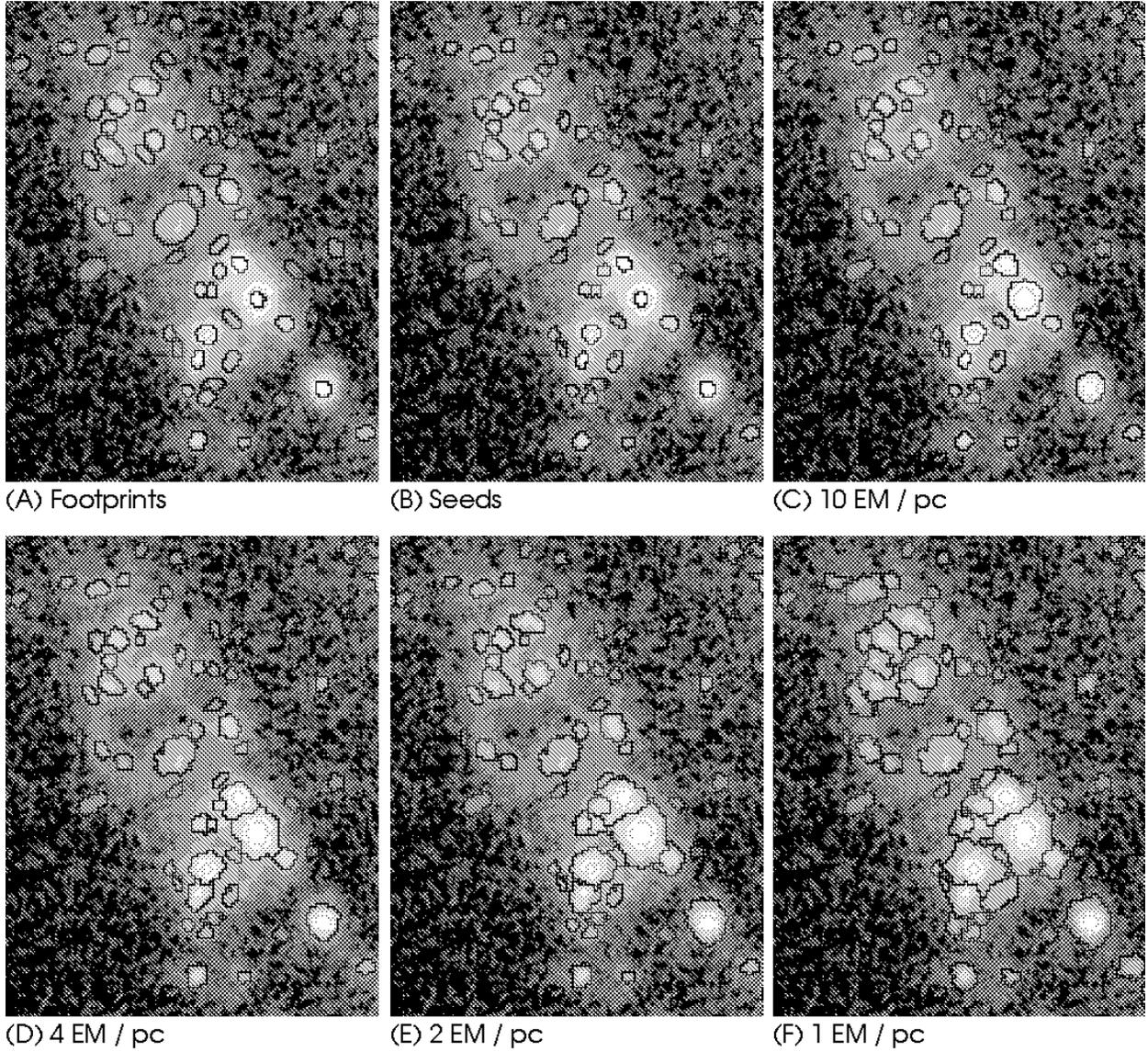}}
\addtocontents{lof}{\setlength{\baselineskip}{\singlespace}
\vspace{\baselineskip}}
\caption[Small continuum-subtracted H$\alpha$
  subsection of M51, shown at various stages of the {\bf{\rm H{\small
  II}{\it phot}}} procedure]{Small continuum-subtracted H$\alpha$
  subsection of M51, shown at various stages of the {\bf{\rm H{\small
  II}{\it phot}}} procedure.  In particular, panels (a) and (b)
  indicate the footprint and seed boundaries, while panels (c-f) show
  the gradual growth of detected \ion{H}{2} regions.  Panel (c) represents
  growth to a terminal surface brightness slope of 10 EM/pc.  Panel
  (d) indicates the end state for growth to 4 EM/pc.  The extent of
  all \ion{H}{2} regions in panel (e) was determined by growing until we
  reach 2 EM/pc, just before our nominal stopping point when the
  surface brightness profile of each region has flattened to at least
  1.5 EM/pc.  See Fig.~\ref{fig:M51.grow} for the nominal result, showing all of M51.
  Finally, panel (f) shows the resulting boundaries when growth is
  continued to 1 EM/pc in an attempt to recover any locally
  concentrated DIG possibly associated with each classical \ion{H}{2} region.}
\label{fig:iterative.growth}
\addtocontents{lof}{\setlength{\baselineskip}{\doublespace}}
\end{figure}

In the end, we elected to permit an array of different stopping points
ranging from very little growth to nearly the generous ``flat result''
described above.  This amounted to adopting a series of cutoffs in
terminal surface brightness slope, [10, 4, 2, 1.5, and 1] EM/pc, then
running the growth procedure from beginning to end for each.  Notice
that the specific cutoff values given here are only appropriate if the
calibrated narrowband data are expressed in the conventional units of
EM, cm$^{-6}$~pc, and must be rescaled for any other case.  In
section~\ref{sec:HIIphotcode.HIILF}, we show that this approach allows
us to directly address systematic uncertainty in \ion{H}{2} region fluxes
(and the resulting luminosity function) associated with our decision
to stop growth at a given point.  Figs.~\ref{fig:HIIphot.demo}c
and~\ref{fig:HIIphot.demo}e present images of the M51 subsection with
\ion{H}{2} region boundaries marked for two different values of the
terminal surface brightness slope.  In Fig.~\ref{fig:HIIphot.demo}c
growth has stopped at a slope of 10 EM/pc, leaving a substantial
fraction of diffuse emission possibly associated with discrete \ion{H}{2}
regions remaining outside the {\bf{\rm H{\small II}{\it phot}}}
boundaries.  Fig.~\ref{fig:HIIphot.demo}e shows the result for growth
continuing until surface brightness profiles flatten to 1 EM/pc.
Notice how isolated regions do eventually stop growing on their own,
while the crowded central area has effectively been subdivided into
numerous chunks (each plausibly associated with an embedded ionizing
source).  The behavior of the iterative growth procedure is further
illustrated in Fig.~\ref{fig:iterative.growth}, which shows another
section of M51 at several stages of growth.  For reference, all the
results presented in this paper are based on a terminal profile slope
of 1.5 EM/pc, midway between the degree of growth shown in panels (e)
and (f) of Fig.~\ref{fig:iterative.growth}.

One important advantage of adopting terminal surface brightness slopes
(specified in physical units) is that our procedure is relatively
robust to changes in signal-to-noise.  One can think of other criteria
for stopping growth that are not as reliable.  For instance, we
initially tried to quantify the ``stopping point'' for \ion{H}{2} region
growth in terms of various critical multiples of the formal error in
the dimensionless surface brightness slope.  This procedure appeared
promising when analyzing our basic dataset, but was found to introduce
substantial bias in the definition of \ion{H}{2} region boundaries during
experiments in which the signal-to-noise was globally reduced by
factors ranging from 2 through 5.  In short, as the test images were
made noisier, growth stopped progressively sooner despite the fact
that the underlying observed surface brightness profiles were no
different.  Our adopted procedure is substantially more well-behaved
under these circumstances and generates luminosity functions which are
statistically indistinguishable at all but the lowest luminosities.
Sadly, loss of low luminosity sources is unavoidable with degraded
signal-to-noise no matter how regions are grown.

\subsection{Correction for underlying emission}
\label{sec:HIIphotcode.diffuse.correction}

For \ion{H}{2} regions embedded in a background of diffuse ionized gas it
is important to accurately estimate the DIG flux contribution to the
observed counts within a region's boundary.  In past studies, most
authors have gauged the background contribution by interactively
selecting one or more positions near each \ion{H}{2} region they thought to
be representative of the level underlying the source.  Our method
works as follows: (1) after final region boundaries are available, we
define as ``background'' pixels all those unclaimed pixels within a
projected distance of 250~pc from the boundary of an \ion{H}{2} region, (2)
next we select a uniformly-spaced set of ``control points'' to
represent these background data, only accepting those which are at
least 75\% surrounded by other background pixels within a circular
domain of 250~pc diameter, (3) we then compute the median value of all
background pixels within the domain of each control point, and (4) we
finally compute a surface fit to these median values.  Our
surface-fitting procedure generates a low-order solution on small
scales by interpolating between the 3 nearest control points at every
position, but in a global sense the product is a very high-order
surface.  The result is essentially an image of the diffuse emission
present in the original data and therefore represents an excellent
means of quantifying the diffuse fraction in galaxies
(e.g. \citet[]{hoopesetal96}).
Note that we compute a different surface-fit for each requested
version of the region boundaries, as the degree of iterative growth
will influence pixel membership in background annuli and therefore the
estimated background level for each emission line source.
Figs.~\ref{fig:HIIphot.demo}d and
\ref{fig:HIIphot.demo}f show the diffuse background for the growth states illustrated in
Figs.~\ref{fig:HIIphot.demo}c and \ref{fig:HIIphot.demo}e,
respectively.  Note how the level of diffuse emission is estimated to
be substantially lower in the second case.

\subsection{{\bf{\rm H{\small II}{\it phot}}} data products}
\label{sec:HIIphotcode.data.products}

The output of {\bf{\rm H{\small II}{\it phot}}} consists of several
images and one catalog detailing properties of all detected regions.
The catalog tabulates the following quantities among others: ID\#,
right ascension, declination, pixel position, number of pixels
contained by the region, effective FWHM, major axis FWHM, axial ratio,
position angle, total flux after correction for background emission,
$1\sigma$ uncertainty in total flux after correction, and the peak
surface brightness inside the region.  See
Section~\ref{sec:HIIphotcode.flux.and.SN} for a description of how
total corrected flux and its error are calculated.  Note that right
ascension, declination, pixel position, and FWHM values refer to the
best-fitting empirical model associated with each region, so from
before region growth.

The images produced by {\bf{\rm H{\small II}{\it phot}}} include: (1)
a copy of the continuum-subtracted line image with the various ``after
growth'' boundaries marked, (2) several versions of the background
surface fit (corresponding to different levels of growth), and (3)
integer maps delineating the position and extent of each footprint,
seed, and grown region.  These integer maps can be used for
supplementary analysis if identically gridded images at different
wavebands are available.  Among the most obvious applications are
computation of line ratios or equivalent width for emission line
objects.  Finally, the {\bf{\rm H{\small II}{\it phot}}} user has the
option of dumping postage stamp collages depicting each source in the
catalog.
 
\subsection{Flux determination and signal-to-noise in {\bf{\rm H{\small II}{\it phot}}}}
\label{sec:HIIphotcode.flux.and.SN}

The background-corrected emission line flux of an \ion{H}{2} region is
computed using the continuum image ($C$), the line+continuum image
($L$), and our {\bf{\rm H{\small II}{\it phot}}} surface fit to
diffuse background emission remaining in the continuum-subtracted line
image after growth of sources ($D$).  In this derivation we assume
that $C$, $L$, and $D$ remain in ADUs.  Additionally, we require that
no sky background has been subtracted from either $C$ or $L$.  This is
essential if photon noise is to be properly modeled during estimation
of signal-to-noise.  For region $i$ (composed of pixels $j =
{1~...~n_i}$) the background corrected emission line flux, $F_i$, is
calculated as:

\begin{equation}
\label{flux}
F_i = \sum ( L_{ij} - S_i C_{ij}) - \sum D_{ij},
\end{equation}

\noindent 
where $S_i$ is the continuum scaling factor appropriate for region
$i$.  In practice we hold $S_i$ constant for all regions.  The formal
$1\sigma$ uncertainty, $\delta F_i$, associated with $F_i$ is given by
the quadratic sum of standard deviations associated with individual
terms of Eq.~\ref{flux}:

\begin{equation}
\label{delta.flux}
\delta F_i = \frac {\sum ( {\delta L_{ij}}^2 + {S_i}^2 {\delta C_{ij}}^2 + {\delta D_{ij}}^2 )^{1/2}}{g}
\end{equation}

As written here, the units of $F_i$ and $\delta F_i$ are ADUs, while
$g$ is the gain in terms of electrons per ADU.  To convert into
physically meaningful units we multiply by an appropriate calibration
factor.  We assume that $C$ and $L$ are both essentially sky-noise
limited, implying the following relations:

\begin{equation}
\label{sigmaLsky}
\delta L_{sky} = \frac{1}{g} \left( \frac{g X_L L_{sky}}{n_L} \right)^{1/2}, {\rm and}
\end{equation}

\begin{equation}
\label{sigmaCsky}
\delta C_{sky} = \frac{1}{g} \left( \frac{g X_C C_{sky}}{n_C} \right)^{1/2}.
\end{equation}

\noindent
In these expressions, $n_L$ and $n_C$ are the number of images
(assumed to have comparable exposure) combined to create $L$ and $C$,
respectively.  Multiplicative factors $X_L$ and $X_C$ have values near
unity or slightly higher in order to account for the possibility that
read-noise may still make a small contribution to the noise budget in
the {\bf{\rm H{\small II}{\it phot}}} sky region.  $\delta L_{sky}, \delta C_{sky}, L_{sky},$
and $C_{sky}$ are each measured within the sky region of the input
images, implying appropriate values for $X_L$ and $X_C$.  This
information then constrains a photon noise model for the data, as we
can fold $X_L$ and $X_C$ together with $g$ to represent an effective
gain ($g_L = g X_L$ and $g_C = g X_C$) for $L$ and $C$.

Next, we estimate the level of noise per pixel in the brighter,
interesting portions of $L$ and $C$.  The relevant equations are:

\begin{equation}
\label{deltaL}
{\delta L_{ij}}^2 = \frac{g_L L_{ij}}{n_L}~e^{-}, {\rm and}
\end{equation}

\begin{equation}
\label{deltaC}
{\delta C_{ij}}^2 = \frac{g_C C_{ij}}{n_C}~e^{-}.
\end{equation}

Eqns.~\ref{deltaL} and \ref{deltaC} specify most of the terms in
Eq.~\ref{delta.flux}.  Variance in the diffuse background level,
${\delta D_{ij}}^2$, is determined empirically on the basis of the
measured standard deviation near control points used during the
background fitting process.  Note that we convert such measurements to
electrons before computation of $\delta F_i$.

Two comments must be put forward at this point.  Our assumption of
sky-noise limited imagery is a conservative approach.  By adopting
Eqs.~\ref{sigmaLsky}--\ref{deltaC}, we guarantee that $\delta L_{ij}$ and
$\delta C_{ij}$ will always be predicted accurately or overestimated.
If readnoise contributes substantially to the standard deviation of
pixel values in the user-selected sky region, the measured values of
$g_L$ and $g_C$ insure that it will contribute an identical fraction
of the estimated noise-budget for pixels that are substantially
brighter (due to observed emission from the object of interest).  In
reality, this is not the case -- detector readnoise is independent of
the observed pixel intensity.  Overprediction of error terms ${\delta
L_{ij}}^2$ and ${\delta C_{ij}}^2$ is the unavoidable consequence.
Furthermore, we argue that our procedure for evaluating the source
signal-to-noise ratio ($F_i /
\delta F_i$), is more accurate than the traditional method based only
on continuum-subtracted data, especially in the limit of bright
continuum emission.  Previous studies have usually gauged the standard
deviation per pixel in one or more selected ``sky areas,'' then added
this term in quadrature based on the number of pixels in a region.
This implies that their estimated signal-to-noise is independent of
the original observed datavalues.  Identical sources located in
various positions on top of a variable background of (continuum or
line) emission will be assigned identical signal-to-noise, even though
the true uncertainty increases for sources embedded in a bright
background.

\section{Narrowband observations of M51}
\label{sec:HIIphotcode.M51.obs}

We obtained narrowband images of the M51/NGC5195 system as part of a
separate project concerning diffuse ionized gas (DIG) in spiral
galaxies (\citet[]{greenawalt98}).  These data were obtained in 1992
March using the No.--1 0.9 m telescope at Kitt Peak.  Three H$\alpha$
+ [\ion{N}{2}] and two [\ion{S}{2}] images of 20 min integration were recorded in
addition to a set of offband continuum exposures.  The bandpass of
each filter was approximately 75\AA.  Complete details concerning our
observations and image reduction can be found in
\citet[]{greenawaltetal98}.  For the present analysis a
slightly different flux calibration was used.  The
continuum-subtracted image originally presented in
Greenawalt et al. was calibrated by comparison with
the photometric data of \citet[]{vanderhulstetal88}.
An identical procedure was employed by \citet[]{rand92} in a
detailed study of 616 M51 \ion{H}{2} regions.  Because we sought to compare
our photometry directly with Rand's, we bootstrapped to
his flux scale using a sample of 10 bright, moderately isolated \ion{H}{2}
regions.  The magnitude of this recalibration amounted only to
$\sim$10\%, most likely attributable to the use of different regions
by Rand and Greenawalt.

The 1$\sigma$ noise of our continuum-subtracted line image is at an
H$\alpha$ emission measure (EM) of 9.9~pc~cm$^{-6}$ at $1.8\arcsec$ FWHM
resolution.  This corresponds to a surface brightness of
$2.0\times10^{-17}$~erg cm$^{-2}$ s$^{-1}$ arcsec$^{-2}$, or 3.5
Rayleighs.  Our noise implies a limiting (5$\sigma$) point source flux
of $3.6\times10^{-16}$~erg cm$^{-2}$ s$^{-1}$, or equivalently an
H$\alpha$ luminosity of $3.9\times10^{36}$~erg s$^{-1}$, neglecting
any background confusion.  For this calculation and the analysis
below, we have assumed a distance of 9.6 Mpc to M51
(\citet[]{sandagetammann74}).  At this distance, the $1.8\arcsec$~seeing
prevailing during our observations corresponds to a linear resolution
of 84~pc.  No correction was attempted for extinction, in order to
facilitate comparison with earlier \ion{H}{2} region surveys.  In reality,
\citet[]{vanderhulstetal88} report on extinction toward a large sample
of M51's most luminous \ion{H}{2} regions, finding about 2 visual
magnitudes in most cases.  This should be kept in mind when
interpreting our results.
 
\section{Results}
\label{sec:HIIphotcode.results}

\begin{figure}
\centering
\resizebox{0.95\hsize}{!}{\includegraphics{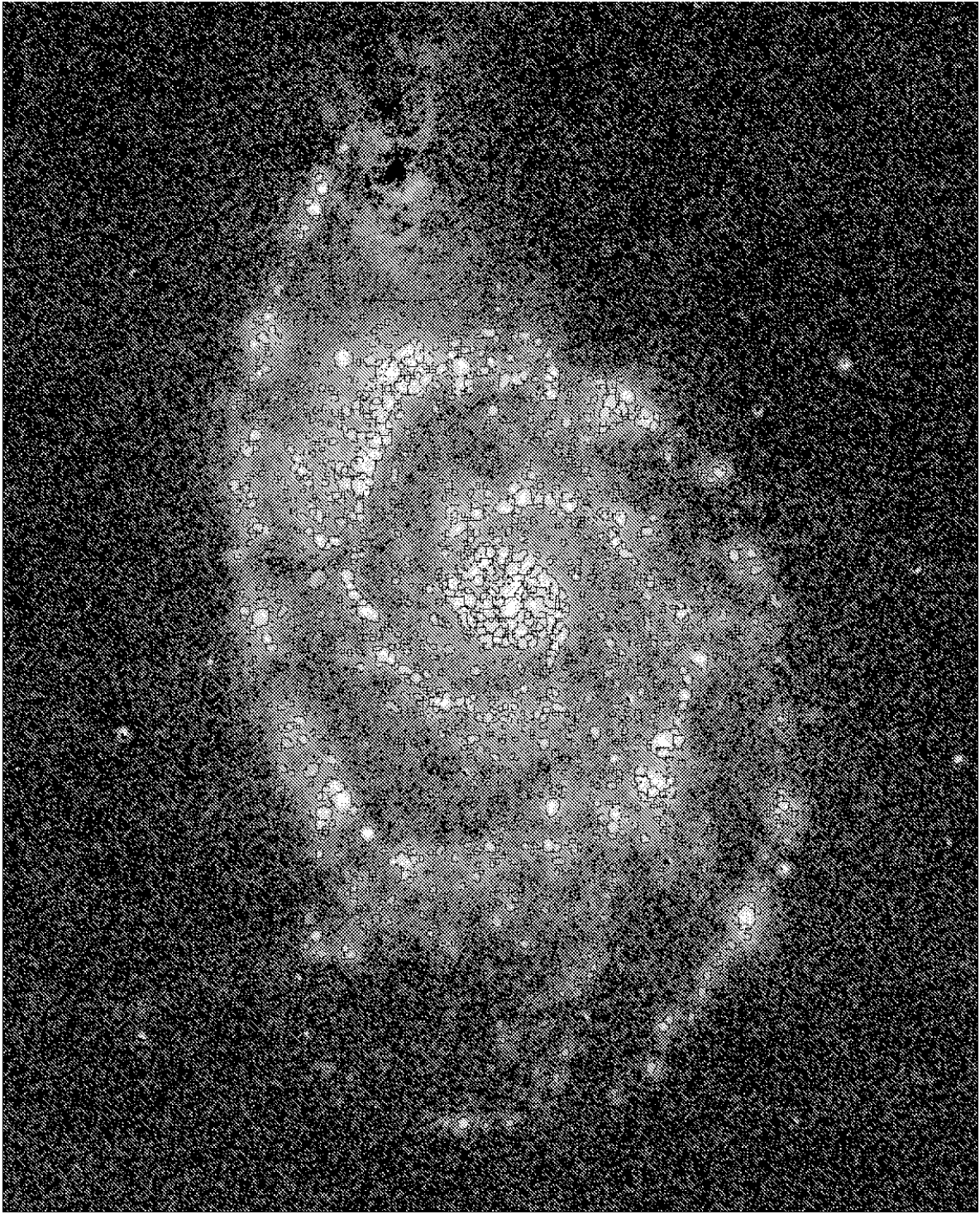}}
\addtocontents{lof}{\setlength{\baselineskip}{\singlespace}
\vspace{\baselineskip}}
\caption[Continuum-subtracted H$\alpha$ image of M51,
  with 1.5 EM/pc {\bf{\rm H{\small II}{\it phot}}} boundaries
  indicated]{Continuum-subtracted H$\alpha$ image of M51,
  with 1.5 EM/pc {\bf{\rm H{\small II}{\it phot}}} boundaries
  indicated.  The image has been logarithmically scaled in order to
  preserve useful contrast over the entire field.  1618 \ion{H}{2} regions
  are marked, having luminosities in the range log~L$_{H\alpha} \sim 36
  - 40$.  Of these, a total of 1229 sources have estimated H$\alpha$
  luminosities determined with at least 20\% accuracy.  Stellar
  residuals have not been blanked out, demonstrating the ability of
  {\bf{\rm H{\small II}{\it phot}}} to ignore such image defects.}
\label{fig:M51.grow}
\addtocontents{lof}{\setlength{\baselineskip}{\doublespace}}
\end{figure}

We detected 1618 \ion{H}{2} regions in the field of view of our
observations, excluding sources located in M51's interacting companion
NGC5195.  Of the total sample, 1229 regions were classified as
``photometric-quality'' detections having ${\frac{S}{N}}_{final} \ge
5$.  Only these 1229 \ion{H}{2} regions have been considered in the
analysis described below.  Fig.~\ref{fig:M51.grow} displays our
continuum-subtracted H$\alpha$ image, with all source boundaries
marked.  Note that the extent of each \ion{H}{2} region has been determined
using a terminal surface brightness slope of 1.5 EM/pc.  All results
in the rest of the paper correspond to this choice, unless stated
otherwise.  The image has been logarithmically scaled in order to
preserve contrast over a wide range in surface brightness.  Notice the
hand-drawn loop surrounding NGC~5195.  It indicates the region
specifically excluded from our M51 {\bf{\rm H{\small II}{\it phot}}}
survey.

\subsection{Overall comparison with R92}
\label{sec:HIIphotcode.R92comparison}

In this section we compare our \ion{H}{2} region catalog with the list
compiled by \citet[]{rand92}, hereafter R92, based on visual inspection
and classification.  We highlight the overall correspondence between
our results and those of R92, but also describe variations
attributable to procedural differences.  We include a look at catalog
completeness as a function of luminosity and morphological properties.

The most straightforward comparison between both catalogs is the total
number of detected \ion{H}{2} regions.  Our detection list encompasses the
entire galaxy, even the outer disk and confused nuclear portions not
considered by Rand, suggesting we should find more sources than
previously reported.  However, other competing factors need to be
considered as well:

(1) During the definition of footprints, {\bf{\rm H{\small II}{\it
      phot}}} considers if a set of neighboring peaks is best
      described as a collection of individual regions or should be
      grouped into one or more composite aggregations.  R92 always
      classified each neighboring peak as a separate source.

(2) {\bf{\rm H{\small II}{\it phot}}} is perfectly consistent during
the evaluation of marginal detections.  Our estimate of
signal-to-noise for each detection is evaluated on the basis of both
the line+continuum and continuum datasets rather than just the
continuum-subtracted line image (see
Section~\ref{sec:HIIphotcode.flux.and.SN}).

(3) The catalogs were generated using images having different
intrinsic sensitivity.  The 1$\sigma$ noise in our data was
$2.0\times10^{-17}$~erg cm$^{-2}$ s$^{-1}$ arcsec$^{-2}$, whereas
Rand's imagery went down to $\sim 1\times10^{-17}$~erg cm$^{-2}$
s$^{-1}$ arcsec$^{-2}$ (EM = 5~pc~cm$^{-6}$), when evaluated at
comparable spatial resolution.

(4) Contamination by emission-line objects other than traditional
\ion{H}{2} regions is a concern in our catalog.  As a fundamental part of
the R92 source selection process, each tentative detection was
individually checked in a number of ways.  Rand demanded that every
source be centrally peaked, have a limited degree of circular
symmetry, and possess a peak flux exceeding the background by more
than 50\%.  {\bf{\rm H{\small II}{\it phot}}} does not use any of
these criteria.  This means that our procedure is much more likely to
result in a catalog containing ``objects'' such as localized
enhancements in the widespread curtain of DIG, whether they be ionized
on the spot by an embedded OB field star or by leakage from an \ion{H}{2}
region in a neighboring part of the galaxy, possibly hundreds of pc
away.  We elected to accept detections of this sort for three reasons:
(1) they are intrinsically interesting, (2) they have little effect on
the derived slope of the \ion{H}{2} region luminosity function, and (3)
eliminating such detections from the catalog would either require
human intervention or an extra {\it a priori} constraint on the
properties of detected \ion{H}{2} regions.  Note that contamination by
planetary nebulae is equally unlikely in the R92 and {\bf{\rm H{\small
II}{\it phot}}} catalogs, since they would be too faint at a distance
of 9.6~Mpc.  \citet[]{vassiliadisetal92} show that in the Magellanic
Clouds there are no planetary nebulae with log L$_{H\alpha} > 36$,
suggesting that both \ion{H}{2} region catalogs are probably uncontaminated
by PNe.

In the range of galactocentric radius examined by Rand (1~kpc $< R_{g}
<$ 15~kpc), he reported 616 individual \ion{H}{2} regions.  {\bf{\rm
H{\small II}{\it phot}}} detected 1184 \ion{H}{2} complexes with fluxes
$\geq 5\sigma$ in the same area.  Some of these objects are composed
of multiple components.  Although the total number of regions
tabulated by {\bf{\rm H{\small II}{\it phot}}} is more than reported
by R92, we find that the agreement is substantially better in the
range log L$_{H\alpha} > 38.5$.  Rand detected 67 \ion{H}{2} regions of
this luminosity class, whereas we have 80.

The astute reader might ask how many of the regions detected by
{\bf{\rm H{\small II}{\it phot}}} are exactly the same sources
described by R92.  We explicitly checked this issue, finding that
{\bf{\rm H{\small II}{\it phot}}} misses only 2 of the 616 regions of
R92.  At the position of these two sources, we inspected our data and
found no evidence for a significant detection.  Note that our
assessment of correspondence between the R92 source list and the
{\bf{\rm H{\small II}{\it phot}}} catalog was completed by way of manual inspection.  During
this process we demanded not only positional agreement, but also
similar size between detections considered as being one in the same.
It is important to note that due to the different methods of
photometry it is unlikely that most sources had exactly the same
effective boundary.

\begin{figure}
\centering
\resizebox{\hsize}{!}{\includegraphics{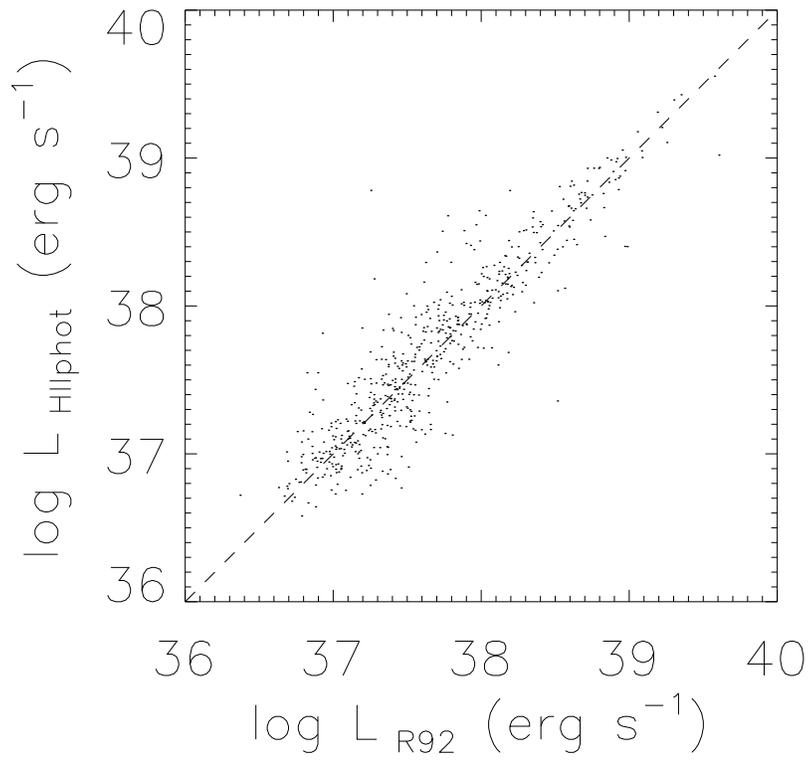}}
\addtocontents{lof}{\setlength{\baselineskip}{\singlespace}
\vspace{\baselineskip}}
\caption[Comparison of measured luminosities
  for 614 sources in common between our catalog and R92]{Comparison of measured luminosities
  for 614 sources in common between our catalog and R92.  The dashed
  line indicates the anticipated unit slope in the case of a perfect
  correlation.  There is substantial scatter about this line, but very
  little systematic deviation is apparent.}
\label{fig:logL.vs.logL}
\addtocontents{lof}{\setlength{\baselineskip}{\doublespace}}
\end{figure}

\begin{figure}
\centering
\resizebox{\hsize}{!}{\includegraphics{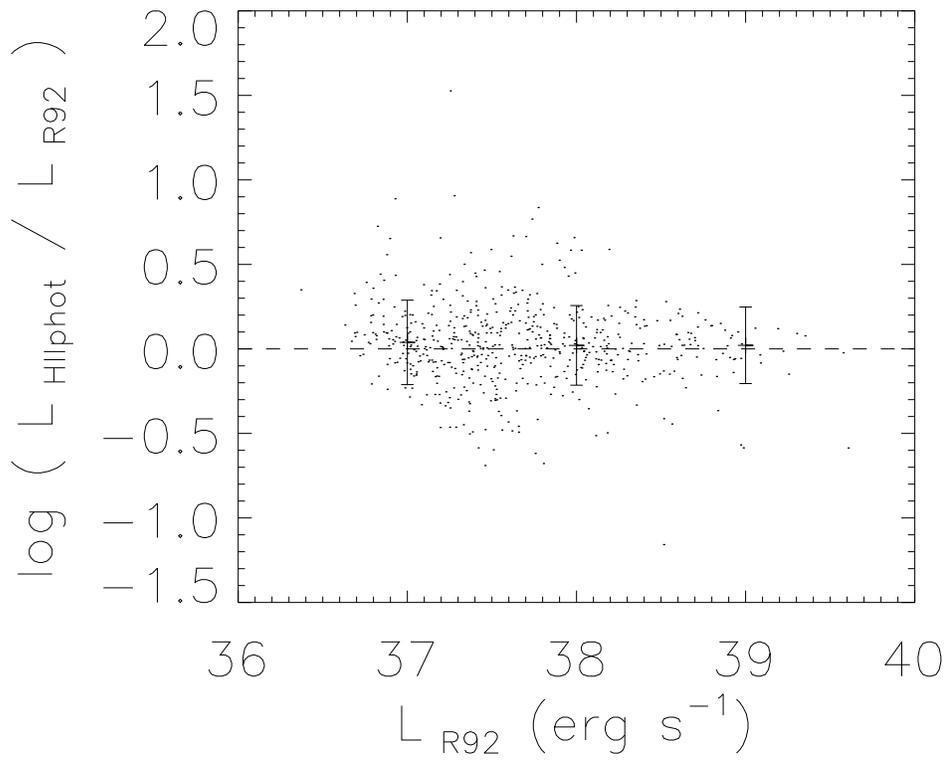}}
\caption[Log ( L$_{{\bf{\rm
        H{\small II}{\it phot}}}}$ / L$_{R92}$ ) versus log L$_{R92}$]{Log ( L$_{{\bf{\rm
        H{\small II}{\it phot}}}}$ / L$_{R92}$ ) versus log L$_{R92}$.
  In this graph, a slight tendency for {\bf{\rm H{\small II}{\it
        phot}}} to measure systematically higher fluxes for the
  faintest regions with respect to the measurements of
  R92 is apparent.  The error bars represent $\pm$
  1$\sigma$, evaluated in bins of width 1.0 dex.}
\label{fig:log.LoverL.vs.logL}
\end{figure}

Figures~\ref{fig:logL.vs.logL} and~\ref{fig:log.LoverL.vs.logL}
present a comparison of the measured luminosities for the regions in
common to both our catalog and the R92 source list.
Fig.~\ref{fig:logL.vs.logL} shows that there is a clear correlation
(having slope $\sim 1$) between the luminosities measured by Rand and
{\bf{\rm H{\small II}{\it phot}}}.  A limited number of \ion{H}{2} regions
fall substantially below the main cloud of datapoints.  These sources
most likely represent cases in which {\bf{\rm H{\small II}{\it phot}}}
broke a single R92 detection into one principal \ion{H}{2} region and a
small number of fainter peripherial sources.
Figure~\ref{fig:log.LoverL.vs.logL} more clearly indicates a very
subtle systematic change in the ratio of {\bf{\rm H{\small II}{\it
phot}}}/R92 luminosities as a function of source strength.  We find
that {\bf{\rm H{\small II}{\it phot}}} tends to return slightly higher
flux levels for very faint sources in comparison to the measurements
of R92.  This trend only begins in earnest at luminosities well below
Rand's estimated completeness limit (at L$_{H\alpha} \sim
10^{37.6}$~erg~s$^{-1}$).  As shown in the next section, this
systematic bias and the inherent scatter in
Fig.~\ref{fig:logL.vs.logL} has very little (if any) influence on the
\ion{H}{2} region LF.

So what are the detections ``missed'' by R92?  In the area surveyed by
Rand, we find that $\sim$80\% of the regions picked up by {\bf{\rm
H{\small II}{\it phot}}} (but not listed in R92) have an H${\alpha}$
luminosity less than $10^{37.5}$~erg~s$^{-1}$.  Morphologically it is
clear why most of them were not included in R92 -- often they are
rather diffuse and/or elongated.  Sometimes these faint sources fall
in the bright halo of a more significant \ion{H}{2} region.  In this case,
interactive inspection of our continuum-subtracted line image
typically reveals a relatively compact source superimposed on the
brighter neighbor.

\subsection{The \ion{H}{2} region luminosity function}
\label{sec:HIIphotcode.HIILF}

\begin{figure}
\centering
\resizebox{\hsize}{!}{\includegraphics{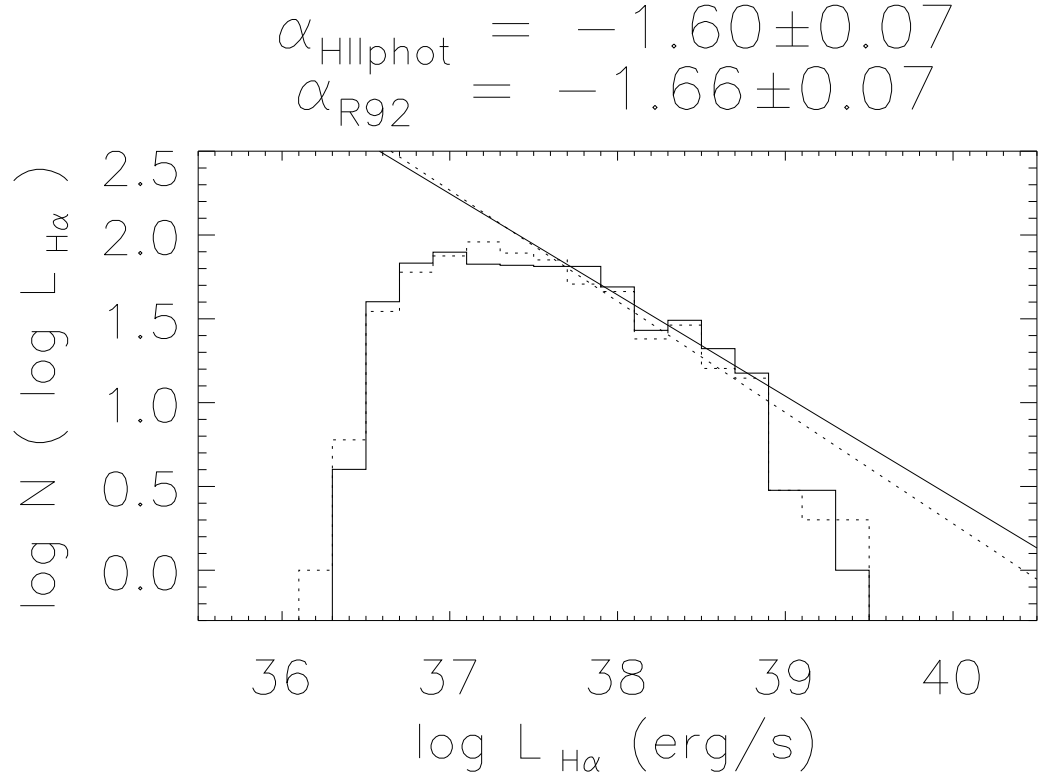}}
\addtocontents{lof}{\setlength{\baselineskip}{\singlespace}
\vspace{\baselineskip}}
\caption[Comparison of the R92 differential \ion{H}{2} luminosity function (for M51) and our equivalent result, only plotting 614 regions for which direct correspondence between catalogs  could be established]{Comparison of the R92 differential \ion{H}{2} luminosity function (for M51) and our equivalent result, only plotting 614 regions for which direct correspondence between catalogs  could be established.  The dotted lines are for R92, whereas solid  lines represent the {\bf{\rm H{\small II}{\it phot}}} data.  In both  cases, we plot the best-fitting power-law function as a straight  line.  Only bins having L$_{H\alpha}$ greater than  10$^{37.6}$~erg~s$^{-1}$ were used in these fits.  The histograms  are essentially identical, illustrating that the {\bf{\rm H{\small II}{\it phot}}} flux measurement procedure does not introduce any detectable bias.}
\label{fig:R92.LFcomp}
\addtocontents{lof}{\setlength{\baselineskip}{\doublespace}}
\end{figure}

Figure~\ref{fig:R92.LFcomp} presents a comparison of the R92
differential luminosity function and our {\bf{\rm H{\small II}{\it
phot}}} result for exactly the same sources.  Above their turnover
points, both luminosity functions can be modeled as power law
distribution (as first suggested by KEH89).  Using only the data for
regions in bins with log~L$_{H\alpha} \ge 37.6$ (those thought to be
complete in R92) and assuming the standard functional form,

\begin{equation}
dN(L) = A L^{\alpha} dL,
\end{equation}

\noindent we find that $\alpha_{\bf{\rm H {\small II}{\it phot}}} =
-1.60 \pm 0.07$, whereas $\alpha_{\rm R92} = -1.66 \pm 0.07$.  (For
these fits we assumed simple counting statistics in order to assign a
variance to each value of the LF.  The weights used during $\chi^2$
minimization were inversely proportional to the variance of each bin,
essentially giving the most influence to bins with the highest number
of detections and reducing the influence of under-populated bins.
This procedure is appropriate as long as no bins suffering from
incompleteness are included in the fit.)  The fact that there is no
significant difference between the results plotted in
Fig.~\ref{fig:R92.LFcomp} indicates that our {\bf{\rm H{\small II}{\it
phot}}} flux measurement technique does not introduce bias into the
observed \ion{H}{2} region luminosity functions.

\begin{figure}
\centering
\resizebox{\hsize}{!}{\includegraphics{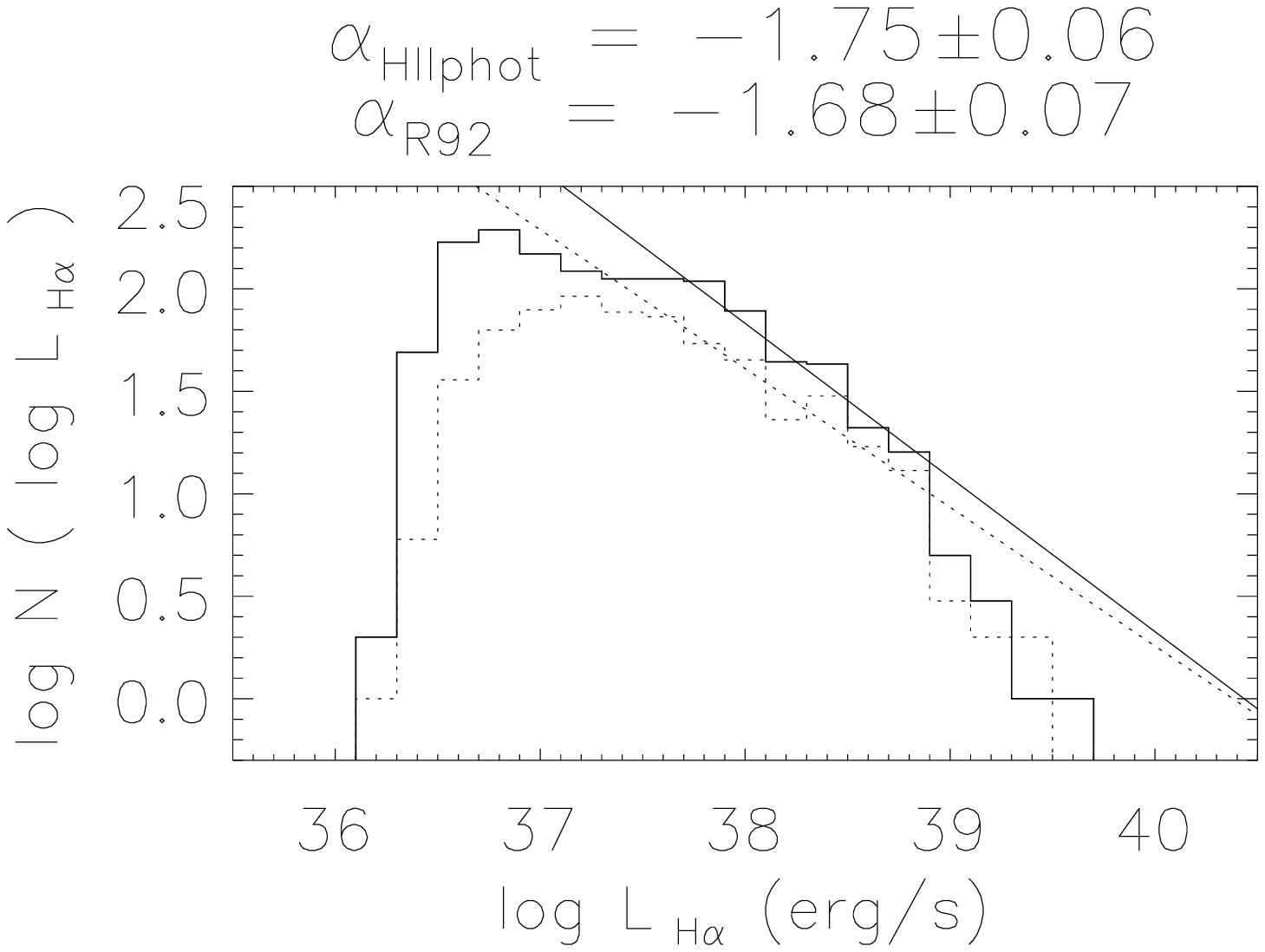}}
\addtocontents{lof}{\setlength{\baselineskip}{\singlespace}
\vspace{\baselineskip}}
\caption[Observed M51 \ion{H}{2} luminosity function,
  obtained using {\bf{\rm H{\small II}{\it phot}}} with the
  establishment of lower limit on signal-to-noise]{Observed M51 \ion{H}{2} luminosity function,
  obtained using {\bf{\rm H{\small II}{\it phot}}} with the
  establishment of lower limit on signal-to-noise
  (${\frac{S}{N}}_{final} \ge 5$).  In total, 1229 regions were
  measured with this degree of accuracy.  The solid line is a weighted
  power-law fit to the bins having L$_{H\alpha}$ greater than the
  inflection at log~L$_{H\alpha}$ = 37.6.  The power-law distribution
  has $\alpha$ = 1.75 $\pm$ 0.06.  Note the prominent break in the LF at
  log~L$_{H\alpha} \sim 38.9$. For comparison we also plot the R92
  result with a dotted line.}
\label{fig:M51.LF}
\addtocontents{lof}{\setlength{\baselineskip}{\doublespace}}
\end{figure}

Restoring the regions ignored for our comparison with R92,
Fig.~\ref{fig:M51.LF} presents an observed LF for all 1229 of the M51
sources detected by {\bf{\rm H{\small II}{\it phot}}} with
$S/N_{final} \geq 5$.  The weighted power-law slope for this
distribution is $\alpha_{\bf{\rm H {\small II}{\it phot}}} = -1.75 \pm
0.06$, including only bins for which log~L$_{H\alpha} \ge 37.6$.
Notice the break in the power-law at a luminosity of 10$^{38.9}$
erg~s$^{-1}$.  Our results confirm that M51 has a Type II LF, as
originally defined by KEH89.  For the purpose of comparison,
Fig.~\ref{fig:M51.LF} also presents the differential luminosity
function from R92 - including the 2 sources not detected by {\bf{\rm
H{\small II}{\it phot}}} (this explains the slight difference with
respect to Fig.~\ref{fig:R92.LFcomp}). Rand estimated that his LF was
complete down to log L$_{H\alpha}$ = 37.6.  In the text below, we
carefully address incompleteness in the {\bf{\rm H{\small II}{\it
phot}}} catalog.  In any case, Fig.~\ref{fig:M51.LF} shows that our
observed LF is marginally steeper than reported in R92.  This probably
reflects the enhanced sensitivity of our procedure at low luminosities
and for relatively diffuse \ion{H}{2} regions.

Note that our \ion{H}{2} LF is subject to a systematic uncertainty
associated with our choice of when to stop growing regions.  In
particular, the observed LF slope varies substantially if growth is
stopped early or allowed to continue until region surface brightness
profiles are more nearly flat.  The LF slope quoted above ($\alpha =
-1.75$) was obtained by growing \ion{H}{2} regions until a terminal surface
brightness slope of 1.5 EM/pc.
If we had instead adopted a 1 EM/pc cutoff, the LF slope would have
been shallower ($\alpha \ge -1.7$).  In the case of minimal growth,
the LF slope approaches $\alpha = -1.9$.  We view this systematic
difference as more of a change in the definition of an \ion{H}{2} region,
rather than uncertainty in our nominal result.  The key point is that
this ``bias'' can be properly addressed in a study of a sample of
spirals by adopting the same procedure for all galaxies.

\subsection{Investigation of incompleteness and blending effects}
\label{sec:HIIphotcode.incompleteness.blending}

It is important to assess incompleteness effects.  We investigated
systematic trends such as the loss of faint, or bright but relatively
diffuse, \ion{H}{2} regions using simulations in which artificial sources
were added to our original images.  These altered data were
subsequently reprocessed using {\bf{\rm H{\small II}{\it phot}}}.

Blending due to limited spatial resolution can induce catalog
incompleteness in crowded environments and flatten the observed LF at
the faint end.  \ion{H}{2} regions tend to be inherently clumpy in terms of
their spatial distribution, especially along spiral arms.  This
implies that the distribution of artificial \ion{H}{2} regions should not
be uniform across an image, but instead that additional \ion{H}{2} regions
should be placed preferentially in areas having recent star formation.
Two ways to achieve this result are described below:

(1) Select a representative subset of detected regions in a galaxy and
permit limited random walks away from actual tabulated positions,
adding an artificial \ion{H}{2} region in each slightly-randomized location.

(2) Use our {\bf{\rm H{\small II}{\it phot}}} surface fit to the
diffuse emission throughout a galaxy as a weighting function for the
probability of placing an artificial \ion{H}{2} region in any particular
spot.

We elected to use the second method, as it provided more flexibility.
For comparison, we also produced simulated images in which we
distributed artificial sources at random.

We sought to reflect the intrinsic morphological diversity of \ion{H}{2}
regions in the prescription employed to generate artificial \ion{H}{2}
regions, rather than just adding unresolved sources of varied
luminosity.  Our incompleteness testing procedure allowed 3 types of
simulated \ion{H}{2} region: (1) small elliptical Gaussians, FWHM$_{eff} =
100$~pc, of varied axial ratio and position angle; (2) large
elliptical Gaussians, FWHM$_{eff} = 200$~pc, of varied axial ratio and
position angle; and (3) background-subtracted copies of actual \ion{H}{2}
regions extracted from the original data (mean FWHM$_{eff}$~=~134~pc),
scaled down to varying flux levels.  In all cases, photon noise was
added to each source before adding it into the line+continuum image
being modified.  Note that all direct image modification was performed
on line+continuum images, since they are the relevant ``observable''
data.  Afterwards, continuum subtraction was performed to compute a
modified continuum-free line image.  {\bf{\rm H{\small II}{\it phot}}}
was run using the modified continuum-subtracted line image, the
modified line+continuum image, and the original continuum image.

Fig.~\ref{fig:M51.LF} suggests that our observed M51 \ion{H}{2} LF might
begin suffering from incompleteness for sources as bright as $\sim
10^{37.0-37.5}$~erg~s$^{-1}$ (just above the turnover point).  We
chose to insert 100 artificial regions of each type at 5 distinct
luminosity values, log$(L_{H\alpha}) = 36.2,~36.6,~37.0,~37.4$, and
$37.8$.  Obviously each combination of source type, luminosity, and
spatial distribution was investigated during a separate trial.  The
goal was to reliably constrain the true (``corrected'') \ion{H}{2} LF down
to log$(L_{H\alpha}) \sim 36.6$.

Because actual (and simulated) \ion{H}{2} regions have a spatial extent
defined by irregular boundaries, one cannot simply inter-compare
center positions for each detection in order to determine if simulated
\ion{H}{2} regions have been recovered.  Each simulated source was assigned
a code indicating whether it was: (1) recovered cleanly, (2) recovered
as a blend, (3) essentially unrecovered, but partially blended with
one other region, (4) essentially unrecovered, but partially blended
with multiple regions, or (5) completely unrecovered.  Sources were
considered to be ``recovered'' (codes 1 and 2) if a single detection
boundary encompassed pixels that contained at least 2/3 of the
inserted region's footprint flux (above the 20\% isophote of the
simulated source), otherwise the source was labeled ``unrecovered''
(codes 3, 4, and 5).  For successfully recovered sources, cleanliness
of recovery was judged by the fraction of total detection flux
contributed by the simulated \ion{H}{2} region.  If a simulated region
contributed at least 50\% of the total flux in a detection, then it
was assumed to be a clean recovery (code 1).  Otherwise, blended
recovery was indicated (code 2).  For unrecovered synthetic sources,
we evaluated the number of neighboring detections claiming at least
one pixel of the unrecovered source footprint.  If no pixels belonging
within the simulated region's 20\% isophote were part of an {\bf{\rm
H{\small II}{\it phot}}} detection, then the artificial source was
considered {\em completely} unrecovered (code 5).  Likewise, if one
and only one {\bf{\rm H{\small II}{\it phot}}} detection claimed a
pixel belonging to the unrecovered source footprint, the synthetic
source was labeled essentially unrecovered, single blend (code 3).  If
more than one detection claimed a synthetic footprint pixel, then code
4 (multiple blend) was indicated.  As a tool for determining the
dependence of recovery statistics and photometric accuracy on
variations in the local environment, we also classified the degree of
crowding in the vicinity of each simulated \ion{H}{2} region.

Before discussing the results of our completeness testing procedure,
we note that the simulations provide a way to quantify the accuracy of
our photometry as a function of luminosity.  Because we know the exact
flux of all simulated regions added to an image, we can determine the
standard deviation of flux measurements for cleanly recovered sources.
We examined the distribution of fractional flux discrepancy, defined
as $( F_{observed} - F_{true} ) / F_{true}$, for each cleanly
recovered artificial source without close neighbors.  We find that the
standard deviation of fractional flux discrepancy increases with
decreasing source luminosity (as expected), ranging from 0.1 for
log$(L_{H\alpha}) = 37.4$ up to 0.3 for log$(L_{H\alpha}) = 36.6$.
Fractional flux discrepancy values were negligible for
log$(L_{H\alpha}) \geq 37.8$.  The measurement scatter is
significantly reduced for small sources.  Furthermore, the median
value of fractional flux discrepancy is very near zero for all but our
faintest artificial sources.

Tables~\ref{tab:weightedsim} and~\ref{tab:uniformsim} present the end
results of our completeness testing procedure for M51.  Specifically,
we have tabulated the percentage of simulated detections falling in
each of the five recovery categories for all region types and
luminosity values.  Table~\ref{tab:weightedsim} indicates the values
for source placement via weighting the distribution of artificial
sources to regions of star formation, while Table~\ref{tab:uniformsim}
shows what was recovered for the uniform source distribution.

As expected the simulations indicate our \ion{H}{2} LF begins to be
substantially incomplete by log~L$_{H\alpha}$ = 37.4 (about
$L_{H\alpha}/ L_{H\alpha, RMS} = 32$).  At this luminosity, nearly one
third of the ``actual'' variety synthetic \ion{H}{2} regions could not be
recovered by {\bf{\rm H{\small II}{\it phot}}} (see
Table~\ref{tab:weightedsim}).  We do find that small sources are
easiest to recover.  Large Gaussians were much more susceptible to
blending with one or more sources.  Actual regions appear to be
intermediate -- harder to recover than 100~pc Gaussians ($1.2\times$
PSF FWHM), but significantly easier than 200~pc Gaussians ($2.4\times$
PSF FWHM).  These statements hold for both the weighted and uniform
source distributions.  Table~\ref{tab:weightedsim}, which shows the
results for our weighted distribution tests, is most appropriate for
the galaxy at large.  However, the uniform distribution recovery
statistics should be used when looking at completeness issues in
uncrowded regions.

The results of our completeness testing procedure allowed us to
perform Monte Carlo simulations designed to gauge systematic bias due
to blending of faint, indistinct regions with brighter sources in
observed luminosity functions.  A separate paper will discuss the
detailed findings of this investigation in a more general context.
However, we were able to show that for the M51 completeness statistics
(presented in Tables~\ref{tab:weightedsim} and~\ref{tab:uniformsim})
the slope of the luminosity function above the low luminosity turnover
was rather insensitive to ``upward contamination'' (see R92)
potentially brought about via blending.

This result is actually somewhat of a coincidence related to the
specific observed power law slope of the M51 LF.  For intrinsically
steeper luminosity functions, having $\alpha \sim -2.0$ for instance,
blending can lead to a shift in the turnover point (to higher L) and
create an artificial hump at slightly higher L (in excess of the true
number of sources per bin).  Shallower LFs than M51 are less
susceptible to blending effects, but suffer severely from
non-detection of low luminosity regions.  For such systems, the
turnover of the LF becomes rather broad and fitting a power law slope
to bins just above the turnover leads to an underestimate of the true
$\alpha$ (that is, we are fooled to think the LF is shallower than it
actually is).  As stated above, the M51 power law slope ($\alpha =
-1.75$) is just shallow enough to avoid severe upward contamination,
but not yet flat enough to substantially change the histogram
character near the turnover point.  Consequently, we conclude the
observed M51 LF slope is rather robust to systematic bias and suspect
that the true (unobservable) LF slope falls within the quoted
uncertainty range for $\alpha$.

\subsection{Distance-related effects on the observed LF}
\label{sec:HIIphotcode.distance.effects}

Systematic application of {\bf{\rm H{\small II}{\it phot}}} to a large
sample of galaxies will be able to address the effects of limited
spatial resolution and sensitivity on observed \ion{H}{2} region luminosity
functions.  Both of these observational characteristics are directly
related to the distance for the object of interest.  We can gauge the
intrinsic bias related to limited resolution and sensitivity by
deriving two LFs for each galaxy, one at the actual distance of the
observed system and another using data which has been degraded to make
the observations appear as if the galaxy was at the distance of our
most-removed system.  Although not really an issue in the present
context, since we are studying a single galaxy which is already
moderately distant (9.6 Mpc), this section has been included to
demonstrate the technique and show that it is rather easy to realize
given the {\bf{\rm H{\small II}{\it phot}}} procedure.

We adopted a conservative procedure for generating image sets
corresponding to the same galaxy at various distances.  Instead of
merely convolving the continuum-subtracted line image, then
regridding, and adding noise (as is typically done), we independently
transform the line+continuum and continuum images, only then creating
the continuum-subtracted result.  This procedure is required to
accurately keep track of the photon statistics associated with
continuum emission underlying \ion{H}{2} regions.  Neglecting this
``hidden'' noise may result in an overestimate of sensitivity when
mimicking the effects of increased distance to a system.

The following step-by-step summary explicitly outlines our procedure:

(1) Select a ``blank sky'' region within the field of view of the
continuum-subtracted line image.

(2) Determine the median level and standard deviation of this sky
region in both the line+continuum and continuum images.

(3) Subtract the respective sky level from the line+continuum and
continuum images.

(4) Determine the total number of galaxy counts in each image.

(5) Convolve with an appropriate Gaussian kernel (having peak of
unity) in order to reduce the spatial resolution in both images.

(6) Regrid the convolved line+continuum and continuum images to a
scale which results in the same PSF as the original data.

(7) Scale down the number of counts in each regridded image to be
consistent with the totals determined in Step 4.  That is, $F_{total,
  new} = F_{total, orig} ({\frac{D_{orig}}{D_{new}}})^2$, where
$D_{orig}$ and $D_{new}$ are the original and increased distances to
the galaxy, respectively.

(8) Add the appropriate sky level back into each image.

(9) Based on the assumption that the original data were
sky-noise-limited, add photon noise according to a model derived from
our blank sky region (see Step~2).  This model insures that the
magnitude of simulated noise is higher in bright parts of the
image.

(10) Using the distant images created in Steps 1-9, perform continuum
subtraction to compute the line only image.

\begin{figure}
\centering
\resizebox{\hsize}{!}{\includegraphics{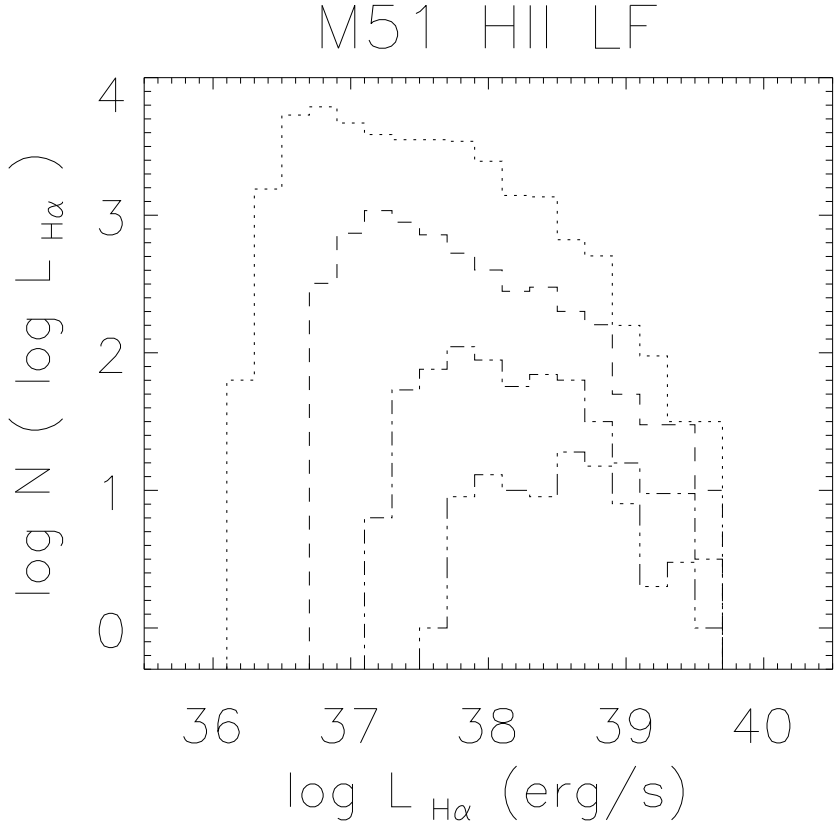}}
\addtocontents{lof}{\setlength{\baselineskip}{\singlespace}
\vspace{\baselineskip}}
\caption[Dependence of the measured \ion{H}{2} region
  luminosity function on systematic effects related to distance,
  including reduced spatial resolution and sensitivity]{Dependence of
  the measured \ion{H}{2} region luminosity function on systematic effects
  related to distance, including reduced spatial resolution and
  sensitivity, is illustrated in this figure.  We plot the M51 LF
  estimated from our original data (at full resolution) with the
  dotted histogram.  After degrading the observations to simulate
  moving M51 to distances of 15, 30, and 45 Mpc, we reprocessed the
  images to derive the apparent LFs.  These results are shown as
  dashed (15 Mpc), dash-dot (30 Mpc), and dash-triple-dot (45 Mpc)
  histograms in this plot.  {\it Histograms for original data, 15 Mpc,
  and 30 Mpc are vertically offset by 1.5, 1.0, and 0.5 dex,
  respectively, in order to preserve clarity.}  Blending and reduced
  sensitivity shift the low luminosity turnover, flatten the general
  slope of the LF, and can boost the brightest detections to
  artificially high apparent luminosities.  Note, however, that even in the presence of these systematic effects the upper end LFs are remarkably consistent in regimes where all datasets are thought to remain complete (log~L$_{\rm H\alpha} \ge 38.6$).}
\label{fig:LF.distance.demo}
\addtocontents{lof}{\setlength{\baselineskip}{\doublespace}}
\end{figure}

For the present demonstration, we generated datasets corresponding to
the appearance of M51 at distances of 15, 30, and 45~Mpc (having PSF
FWHM $\sim$ 130, 260, and 390~pc respectively).
Fig.~\ref{fig:LF.distance.demo} shows a comparison of the \ion{H}{2} LFs
obtained by running {\bf{\rm H{\small II}{\it phot}}} on these
degraded data.  The dotted line traces the actual observed M51 LF,
presented earlier in Fig.~\ref{fig:M51.LF}.

Two effects are rather striking.  The completeness limit at low
luminosities increases in a smooth but dramatic fashion.  Moving from
9.6 Mpc to 15 Mpc, the rapid loss of faint, isolated point sources
takes place and our incompleteness limit (in this case judged by the
LF turnover) rises slightly faster than one might expect according to
the inverse square law.  This effect is mitigated as the galaxy gets
even more distant.  Perhaps blending allows a small fraction of
adjacent weak sources to be recovered as single (brighter) objects.
The second striking effect shown in Fig.~\ref{fig:LF.distance.demo} is
the influence of blending on the slope of the LF.  It is clear that
the LFs tabulated for 30 and 45 Mpc have substantially shallower
power-law slopes than the 9.6 Mpc LF over a limited range of
luminosity.  Indeed, the best-fit LF slope ranges from $\alpha = -1.75
\pm 0.06$ for the original data, to $\alpha = -1.22 \pm 0.08$ for the
case of M51 at 45 Mpc (fitting only sources with log~L$_{H\alpha} >
38.0$).  This effect is brought about by blending of \ion{H}{2} regions (as
spatial resolution is degraded) with some help from catalog
incompleteness.  Blending also explains the increased apparent
luminosity of the brightest \ion{H}{2} regions as the galaxy becomes more
distant, although this effect is not illustrated by
Fig.~\ref{fig:LF.distance.demo} (due to the choice of bin size).

It is worth noting that above a limiting luminosity, the original and
degraded \ion{H}{2} LFs are essentially identical within the errors.  For
this example, the completeness tests of
Section~\ref{sec:HIIphotcode.incompleteness.blending} imply that all
versions of the M51 data in Fig.~\ref{fig:LF.distance.demo} are
complete above log~L$_{H\alpha}$ $\sim$ 38.6.  In the few histogram
bins above this limit, minimal difference between the various LFs is
apparent.

\subsection{Comparison of arm \& interarm regions}
\label{sec:HIIphotcode.arm.interarm}

The results of R92 included a demonstration of changes in \ion{H}{2} region
properties for those sources located in interarm gaps.  We can
classify arm/interarm status based on masking of the diffuse
background image produced by {\bf{\rm H{\small II}{\it phot}}}.  Using
this technique we confirm the difference in LF slope observed by Rand
for arm versus interarm \ion{H}{2} region populations.

A simple way to designate \ion{H}{2} regions as belonging in the arm or
interarm populations relies on masking of the {\bf{\rm H{\small
II}{\it phot}}} surface fit to the diffuse emission remaining after
definition of \ion{H}{2} region boundaries.  These images typically show
very conspicuous spiral structure.  We experimented with several
isophotal cutoffs to obtain a boundary that closely resembled that of
R92.  Although the present goal was to see if we could develop a
masking technique to efficiently reproduce the classification scheme
of Rand (who carefully subdivided the entire sample of \ion{H}{2} regions
on the basis of spiral arm morphology), a more appropriate
characterization of our new method would be one in which the isophotal
mask is used to segregate regions on the basis of their local star
forming environment.  Under the assumption that more DIG is found in
areas of enhanced recent star formation, the ``arm'' \ion{H}{2} regions
identified by our mask could be thought of as sources that lie within
especially active star forming areas of the galaxy.  In the end, a
cutoff at an H$\alpha$ EM of 30~pc~cm$^{-6}$ worked well in both
contexts for M51 (with $i = 20\arcdeg$).

\begin{figure}
\centering
\resizebox{\hsize}{!}{\includegraphics{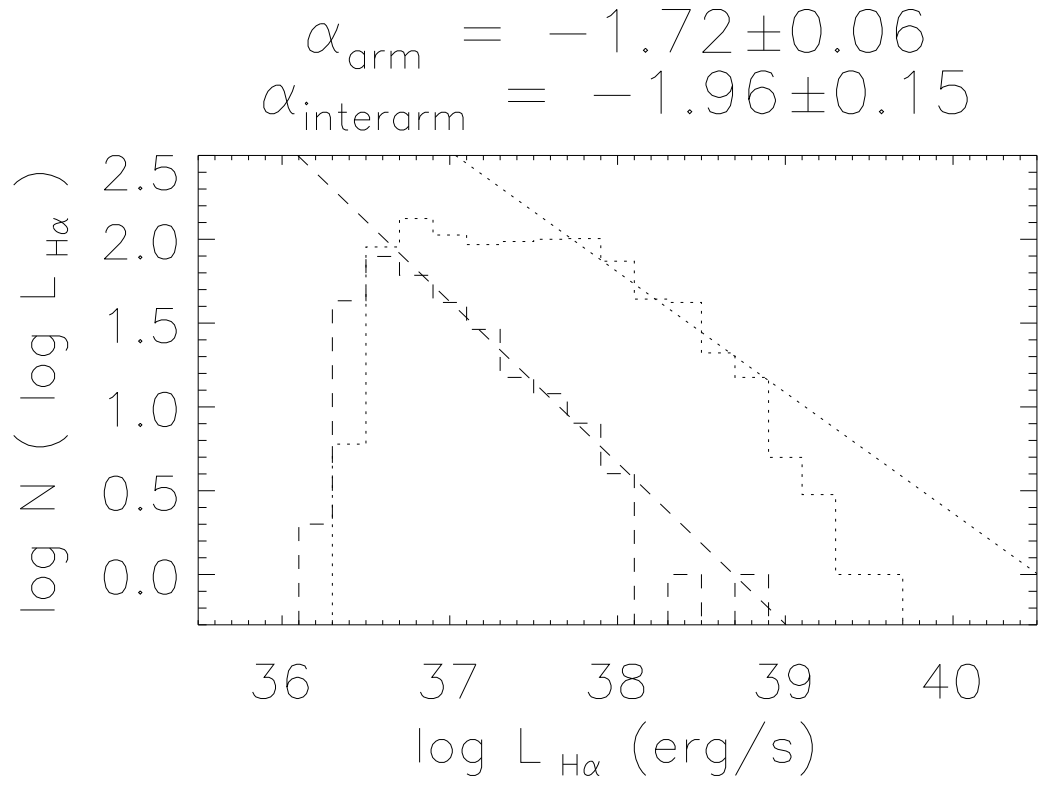}}
\addtocontents{lof}{\setlength{\baselineskip}{\singlespace}
\vspace{\baselineskip}}
\caption[\ion{H}{2} LFs associated with spiral arm/interarm regions as classified on the basis of our EM masking
  technique]{\ion{H}{2} LFs associated with spiral arm/interarm regions as classified on the basis of our EM masking
  technique.  Notice the difference in power law slope for
  arm/interarm regions.}
\label{fig:M51.arm.vs.interarm}
\addtocontents{lof}{\setlength{\baselineskip}{\doublespace}}
\end{figure}

Fig.~\ref{fig:M51.arm.vs.interarm} shows the arm and interarm LFs created using the mask described
above.  The straight lines are weighted power-law fits to the data for
\ion{H}{2} regions brighter than log~L$_{H\alpha}$ = 37.6 and 37.0, for arm
and interarm respectively.  Our simulations of the previous section
suggest that the catalog of interarm sources is complete to lower
luminosities than the general population.  This was the motivation
behind choosing different lower limits for arm and interarm power-law
fits.  We find that there is a difference in slope between the two
populations.  The spiral arm population is best-fit with $\alpha =
-1.72 \pm 0.06$, whereas the interarm regions have a much steeper
power-law slope given by $\alpha = -1.96 \pm 0.15$.  The brightest
\ion{H}{2} regions are found almost exclusively within the spiral arms.
Only two interarm \ion{H}{2} regions in M51 are more luminous than
L$_{H\alpha}$ = 10$^{38.1}$~erg~s$^{-1}$.

\subsection{Correlation between H$\alpha$ luminosity and \ion{H}{2} region size}
\label{sec:HIIphotcode.LHA.vs.PSA}

\begin{figure}
\centering
\resizebox{\hsize}{!}{\includegraphics{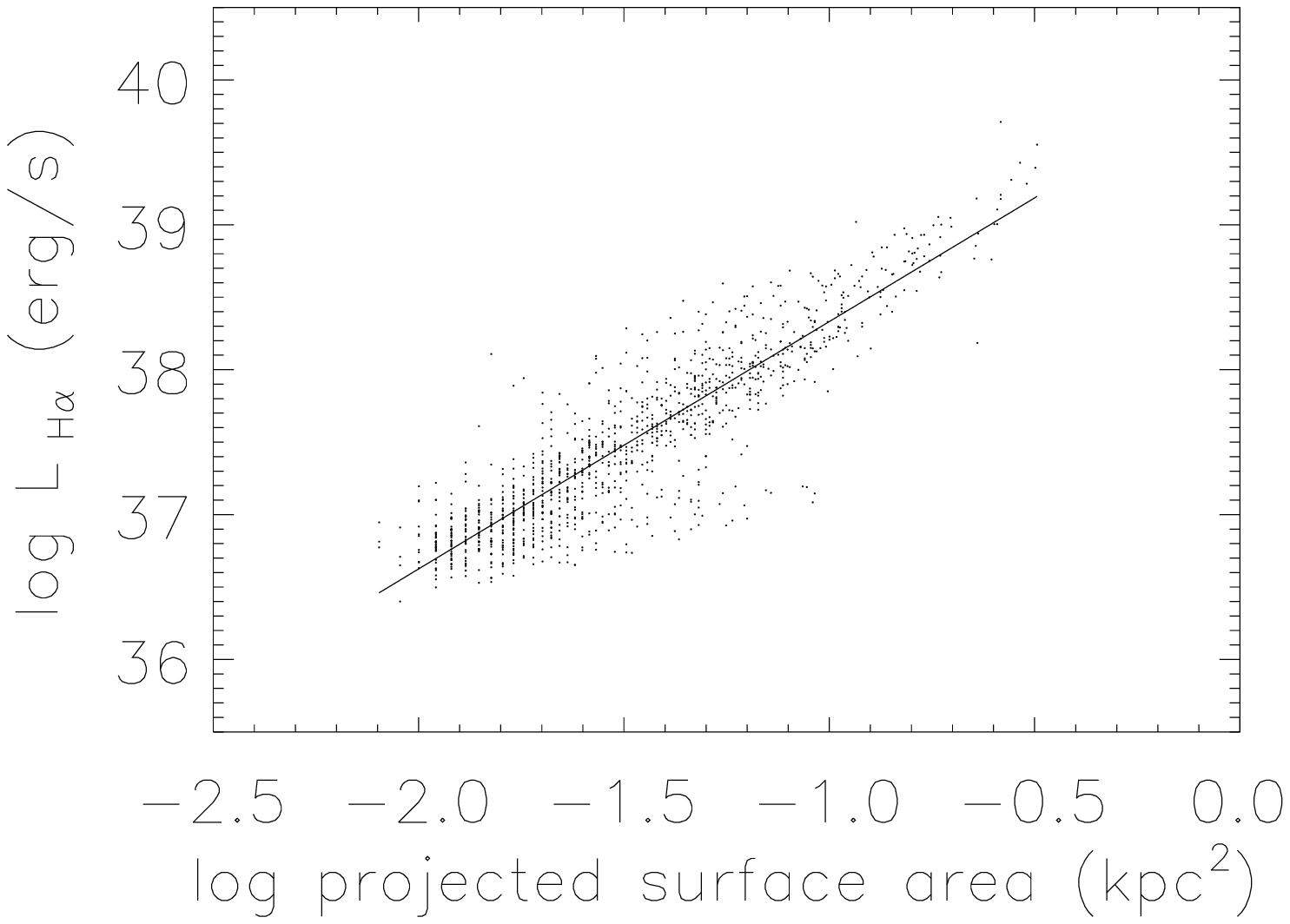}}
\caption[Correlation between log L$_{H\alpha}$ and
  log (projected surface area)]{Correlation between log L$_{H\alpha}$ and
  log (projected surface area) is observed for the \ion{H}{2} regions of
  M51.  The solid line shows a least absolute deviation fit to the
  datapoints and has slope 1.71, significantly higher than what one
  would expect for radiation bounded \ion{H}{2} regions (slope = 1.5).}
\label{fig:logL.vs.logPSA}
\end{figure}

We find that there is a correlation between the H$\alpha$ luminosity
of a region and its projected surface area (PSA).
Fig.~\ref{fig:logL.vs.logPSA} shows a plot of log L$_{H\alpha}$ versus
log PSA.  Our data is best-fit by a line of slope 1.71, substantially
higher than the predicted value of 1.5 for a classical (radiation
bounded) Stromgren sphere of constant density.  The scatter about the
fit is rather large, especially for small \ion{H}{2} regions.  Note that we
have chosen to present the correlation between log L$_{H\alpha}$ and
log PSA, rather than log of \ion{H}{2} region effective radius (r$_{eff}$),
because projected surface area is more directly related to our
observations in the case of sources having irregular shape.  We
suspect that the slightly steeper than expected slope in
Fig.~\ref{fig:logL.vs.logPSA} could be related to clumping within
\ion{H}{2} regions.

\subsection{Characteristics of the DIG}
\label{sec:HIIphotcode.DIGcharacteristics}

The diffuse fraction in spiral galaxies remains of substantial
interest for studies of ISM morphology and energetics.  Defined as the
ratio of DIG H$\alpha$ luminosity to total H$\alpha$ luminosity
(\citet[]{walterbosbraun94}), the diffuse fraction has been estimated in
a number of ways by different authors.  The most common techniques are
based on isophotal masking (e.g.
\citet[]{fergusonetal96}, \citet[]{hoopesetal96}, \citet[]{wangetal99}), although
authors usually disagree on precise methodology.  Classification of
DIG has also been accomplished using explicit identification of
traditional \ion{H}{2} regions (\citet[]{walterbosbraun94}) and using maps of
H$\alpha$ equivalent width (\citet[]{veilleuxetal95}).  It is remarkable
that the results obtained using diverse methods are quite similar,
with a diffuse fraction of $0.4 \pm 0.1$ being common for spiral
galaxies (\citet[]{greenawalt98}).  Nevertheless, the diversity of methods
employed makes it difficult to compare results in a detailed manner
and accurately address relative uncertainties.

There are a few obvious drawbacks to each of the techniques previously
used to estimate the diffuse fraction.  In particular, inspection of
the Fig.~2 in \citet[]{fergusonetal96}, Fig.~4b in
\citet[]{hoopesetal96}, and Fig.~8 in
\citet[]{greenawaltetal98} reveals many instances of
faint but highly localized H$\alpha$ emission being lumped into the
DIG.  These sources could be compact \ion{H}{2} regions or even planetary
nebulae.  Attributing the flux of these faint discrete sources to DIG
tends to artificially boost the diffuse fraction by a small (perhaps
insignificant) amount.  Secondly, several authors have pointed out
that the total DIG luminosity should receive a contribution from
locations in which \ion{H}{2} regions are projected onto a slowly varying,
diffuse background.  The most commonly adopted solution is to assume
that pixels occupied by an \ion{H}{2} region each contribute the mean DIG
intensity when totaling up DIG.  This is undoubtedly an underestimate,
as \ion{H}{2} regions often have prominent DIG haloes, implying that the
DIG superimposed on \ion{H}{2} regions will typically be brighter than
average.  {\bf{\rm H{\small II}{\it phot}}} addresses both of these
problems, because it first measures flux associated with all discrete
emission line sources and then individually estimates a background
level for each region.

We calculate the diffuse fraction by independently totaling: (1)
F$_{HII}$, the background-corrected flux associated with all detected
\ion{H}{2} regions (except those with ${\frac{S}{N}}_{final} < 5$) , and (2)
F$_{tot}$, the flux of the entire image.  The diffuse fraction is then
given by (F$_{tot}-$F$_{HII}$)/F$_{tot}$.  This is the method of
\citet[]{walterbosbraun94}, but accomplished in a repeatable automated
manner.  By computing the diffuse fraction for various requested
stopping-points during the iterative growth process, we can accurately
constrain the diffuse fraction and also place an upper limit on the
amount of DIG ionized in the field, apparently unrelated to classical
\ion{H}{2} regions.

Using region boundaries established by our nominal 1.5 EM/pc terminal
surface brightness slope, we find that the diffuse fraction for M51 is
$0.45 \pm 0.01$.  The uncertainty quoted here only accounts for the
possibility of variation in the sky background.  Other uncertainties
such as those associated with continuum subtraction and growth
termination criteria will also play a role, as will flat-fielding
errors.  In fact, as described below, these factors may actually
dominate the diffuse fraction uncertainty.

\begin{figure}
\centering
\resizebox{0.95\hsize}{!}{\includegraphics{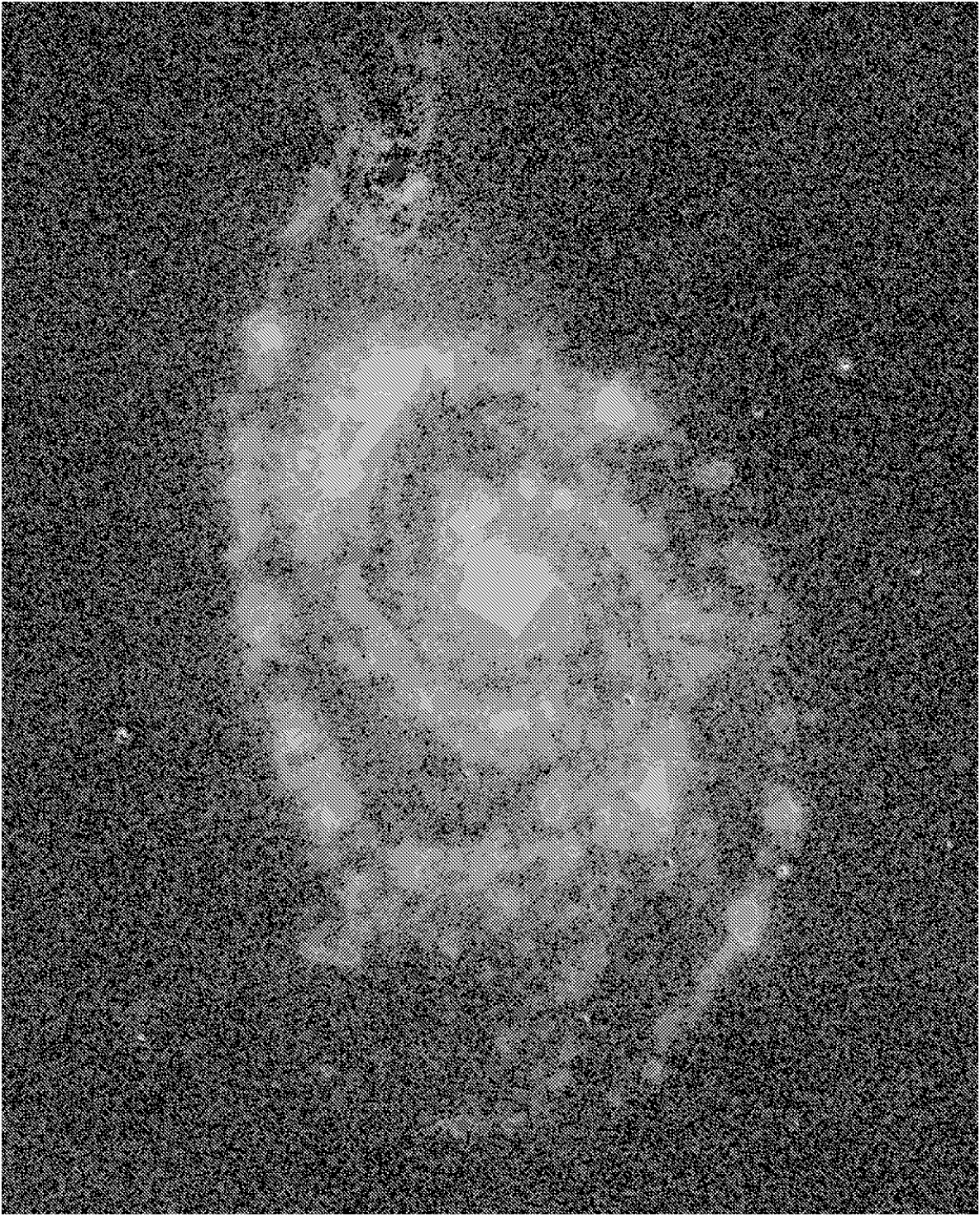}}
\addtocontents{lof}{\setlength{\baselineskip}{\singlespace}
\vspace{\baselineskip}}
\caption[Combination of the original continuum-subtracted image and the {\bf{\rm H{\small II}{\it phot}}}
  surface fit to the diffuse background emission remaining after
  iterative growth]{Combination of the original continuum-subtracted image and the {\bf{\rm H{\small II}{\it phot}}}
  surface fit to the diffuse background emission remaining after
  iterative growth.  The diffuse surface fit has only been shown in regions
  coincident with detected \ion{H}{2} regions.  Note the clear spiral
  structure evident in this logarithmically scaled image.  Images such
  as this are used to estimate the diffuse fraction over the entire
  galactic disk.}
\label{fig:M51.bkgd}
\addtocontents{lof}{\setlength{\baselineskip}{\doublespace}}
\end{figure}

Just over half of the observed H$\alpha$ emission from M51 can be
unambiguously associated with classical \ion{H}{2} regions.
Fig.~\ref{fig:M51.bkgd} presents portions of our {\bf{\rm H{\small
II}{\it phot}}} surface fit to control points located in the diffuse
emission not overlapped by \ion{H}{2} regions.  In this plot, we have only
shown the diffuse background surface fit for pixels covered by an
\ion{H}{2} region - all other areas show the original data.  Panels (d) and
(f) of Fig.~\ref{fig:HIIphot.demo} present the entire smoothly varying
surface fit for comparison.  Note that there is still substantial
spatial correlation between areas of bright DIG and obvious
concentrations of \ion{H}{2} regions.  Indeed, bright rims around a
significant number of \ion{H}{2} regions can be seen in
Fig.~\ref{fig:M51.bkgd}.  Taken together, these facts seem to imply
that we have not yet recovered all the H$\alpha$ emission which is
powered by Lyman continuum photon sources inside traditional \ion{H}{2}
regions.  It is entirely possible that our 1.5 EM/pc growth limit
remains too conservative and should really be lowered if we want to
accurately characterize the massive stars in \ion{H}{2} regions.

Figs.~\ref{fig:HIIphot.demo}e and~\ref{fig:iterative.growth}f
represent the case in which \ion{H}{2} regions were grown to encompass all
apparently associated emission down to the sensitivity limit of the
current data (by adopting a 1 EM/pc stopping point during the
iterative procedure).  Estimating the diffuse fraction with this set
of region boundaries leads us to conclude that {\it at most} 38\% of
M51's H$\alpha$ luminosity might originate via ionization by some
mechanism other than leakage of Lyman continuum photons from \ion{H}{2}
regions.  We do not mean to say that the conventional diffuse fraction
is 0.38, but instead that a substantial fraction of the DIG emission
in M51 cannot be plausibly tied to specific \ion{H}{2} regions with the
current data.  This emission is still somewhat spatially correlated
with the local density of \ion{H}{2} regions, but ionization by ``field''
sources such as OB stars not in associations (Hoopes et al. 1999, in
prep) may be largely responsible.

Our determination of the nominal diffuse fraction for M51 is clearly
subject to systematic uncertainties related to our choice of the
terminal surface brightness slope and uncertainty in the determination
of the scale factor used during continuum-subtraction.  Both of these
can be empirically gauged.  By computing the diffuse fraction
immediately after growth commences (with a 10 EM/pc cutoff, see
Fig.~\ref{fig:iterative.growth}c), we obtain a hard upper limit of
0.68.  At the very least, 32\% of the H$\alpha$ emission from M51 is
contained within the cores of classical \ion{H}{2} regions.  Systematic
changes related to error in continuum-subtraction are easily measured
by producing new versions of the line-only image then recomputing the
background surface-fit and \ion{H}{2} region fluxes.  We generated ``test''
H$\alpha$ images by varying the continuum-subtraction scale factor
$\pm$ 3\% (1 $\sigma$) from our best-guess value.  In the case of 1.5
EM/pc nominal growth boundaries, this resulted in diffuse fractions of
0.40 and 0.49, respectively for increased and decreased continuum
emission.

We note in passing that the diffuse fraction is also potentially
influenced by extinction variations across the face of a galaxy.  The
optical depth towards \ion{H}{2} regions is probably elevated with respect
to field DIG.  Unfortunately, correcting for this systematic error
would be rather difficult even in the case of measured Balmer
decrements, given the unavoidable uncertainty in the geometry of
emitting and absorbing volumes.

\section{Summary}
\label{sec:HIIphotcode.summary}

We have developed a new IDL procedure, which we have designated
{\bf{\rm H{\small II}{\it phot}}}, which is capable of performing
fully-automated, repeatable photometry of \ion{H}{2} regions.  The
procedure can detect and accurately characterize faint sources
embedded in crowded fields, even in the presence of a substantially
inhomogeneous, diffuse H$\alpha$ background.

In this paper we have applied {\bf{\rm H{\small II}{\it phot}}} to the
analysis of the grand-design spiral M51, studied previously by
R92 and KEH89.  Our results are in general agreement
with these authors, although we detect more than twice the number of
\ion{H}{2} regions described by R92.  In total, we find 1229
sources above $5\sigma$ having luminosity greater than about
10$^{36.1}$ erg~s$^{-1}$.  The LF obtained from this catalog of
$5\sigma$ sources is reasonably well fit by a power law distribution
having $\alpha$ = 1.75 $\pm$ 0.06, below a break in the observed
number of regions near log L$_{H\alpha}$ = 38.9.  This break confirms
the earlier classification of M51 as exhibiting a Type II LF.

In the near future, we plan to apply {\bf{\rm H{\small II}{\it phot}}}
for the analysis of an extensive galaxy sample for which high-quality,
sensitive narrowband observations already exist.  The sample will
contain substantially more galaxies than observed by KEH89.  Given the
{\bf{\rm H{\small II}{\it phot}}} code, it will be trivial to
``reobserve'' each of the galaxies at a common distance in order to
inter-compare LFs in the absence of bias associated with different
degrees of blending due to limited resolution and sensitivity.  As a
predecessor to this large study, we present {\bf{\rm H{\small II}{\it
phot}}} results for a smaller sample of 11 spirals in Thilker et
al. (2000, Paper II).  Therein we also develop a procedure for fitting
HII LFs with predictions from population synthesis models of star
cluster formation and evolution.



\acknowledgments
\label{sec:HIIphotcode.acknowledgements}

DAT gratefully acknowledges the support and encouragement of RB and
RW, his dissertation advisors.  DAT further acknowledges the congenial
staff of NFRA for their hospitality during many collaborative trips to
work with RB.  Veronica Fierro has also been of great help to the
authors, finding bugs in our code via repeated trial-and-error
throughout the development stage of {\bf{\rm H{\small II}{\it phot}}}.
DAT has been funded through the NASA Graduate Student Researcher
Program (NGT-51640) and by NSF grant AST9617014 to RAMW.  The {\bf{\rm
H{\small II}{\it phot}}} IDL source code and explanatory documentation
will soon be available by request from DAT.  Contact dthilker@nrao.edu
for details.





\clearpage






\begin{table}
\tiny
\caption[Completeness evaluation results for weighted source distribution]{Completeness evaluation results for weighted source distribution
\label{tab:weightedsim}}
 \begin{center}
 \begin{tabular}{cccccccc} \tableline\\
Source & log $L_{H\alpha}$ & $L_{H\alpha}/ L_{H\alpha, RMS}$  & Recovered & Recovered & Essentially & Essentially & Completely\\
type &  &  & cleanly & as a blend & unrecovered & unrecovered & unrecovered\\
 & & & & & (single blend) & (multiple blend) & \\
 & (erg s$^{-1}$) & & (\%) & (\%) & (\%) & (\%) & (\%) \\
\\
\tableline\\
100 pc Gaussian, $1.2\times$PSF & 37.8 & 79 & 78 & 22 & 0 & 0 & 0 \\
100 pc Gaussian, $1.2\times$PSF & 37.4 & 32 & 67 & 30 & 0 & 3 & 0 \\
100 pc Gaussian, $1.2\times$PSF & 37.0 & 13 & 50 & 43 & 3 & 4 & 0 \\
100 pc Gaussian, $1.2\times$PSF & 36.6 & 5 & 27 & 59 & 2 & 10 & 2 \\
100 pc Gaussian, $1.2\times$PSF & 36.2 & 2 & 4 & 44 & 19 & 19 & 14 \\
\\
\tableline\\
200 pc Gaussian, $2.4\times$PSF & 37.8 & 79 & 43 & 25 & 0 & 32 & 0 \\
200 pc Gaussian, $2.4\times$PSF & 37.4 & 32 & 30 & 26 & 1 & 43 & 0 \\
200 pc Gaussian, $2.4\times$PSF & 37.0 & 13 & 21 & 14 & 5 & 60 & 0 \\
200 pc Gaussian, $2.4\times$PSF & 36.6 & 5 & 2 & 15 & 8 & 66 & 9 \\
200 pc Gaussian, $2.4\times$PSF & 36.2 & 2 & 0 & 18 & 11 & 66 & 5 \\
\\
\tableline\\
Actual region   & 37.8 & 79 & 59 & 21 & 0 & 20 & 0 \\
Actual region   & 37.4 & 32 & 40 & 29 & 0 & 31 & 0 \\
Actual region   & 37.0 & 13 & 34 & 18 & 6 & 42 & 0 \\
Actual region   & 36.6 & 5 & 9 & 31 & 12 & 43 & 5 \\
Actual region   & 36.2 & 2 & 2 & 15 & 22 & 47 & 14 \\
\\
\tableline
\end{tabular}
\end{center}
\end{table}

\clearpage

\begin{table}
\tiny
\caption[Completeness evaluation results for uniform source distribution]{Completeness evaluation results for uniform source distribution
\label{tab:uniformsim}}
 \begin{center}
 \begin{tabular}{cccccccc} \tableline\\
Source & log $L_{H\alpha}$ & $L_{H\alpha}/ L_{H\alpha, RMS}$ & Recovered & Recovered & Essentially & Essentially & Completely\\
type &  & & cleanly & as a blend & unrecovered & unrecovered &
 unrecovered\\
 & & & & & (single blend) & (multiple blend) & \\
 & (erg s$^{-1}$) & & (\%) & (\%) & (\%) & (\%) & (\%) \\
\\
\tableline\\
100 pc Gaussian, $1.2\times$PSF & 37.8 & 79 & 93 & 7 & 0 & 0 & 0 \\
100 pc Gaussian, $1.2\times$PSF & 37.4 & 32 & 82 & 18 & 0 & 0 & 0 \\
100 pc Gaussian, $1.2\times$PSF & 37.0 & 13 & 86 & 13 & 0 & 0 & 1 \\
100 pc Gaussian, $1.2\times$PSF & 36.6 & 5 & 47 & 43 & 3 & 5 & 2 \\
100 pc Gaussian, $1.2\times$PSF & 36.2 & 2 & 8 & 40 & 15 & 13 & 25 \\
\\
\tableline\\
200 pc Gaussian, $2.4\times$PSF & 37.8 & 79 & 67 & 7 & 1 & 25 & 0 \\
200 pc Gaussian, $2.4\times$PSF & 37.4 & 32 & 55 & 10 & 2 & 33 & 0 \\
200 pc Gaussian, $2.4\times$PSF & 37.0 & 13 & 23 & 21 & 11 & 44 & 1 \\
200 pc Gaussian, $2.4\times$PSF & 36.6 & 5 & 8 & 9 & 19 & 54 & 12 \\
200 pc Gaussian, $2.4\times$PSF & 36.2 & 2 &  0 & 13 & 22 & 47 & 21 \\
\\
\tableline\\
Actual region   & 37.8 & 79 &  84 & 8 & 0 & 10 & 0 \\
Actual region   & 37.4 & 32 &  68 & 15 & 0 & 17 & 0 \\
Actual region   & 37.0 & 13 &  46 & 18 & 3 & 32 & 1 \\
Actual region   & 36.6 & 5 &  23 & 16 & 21 & 30 & 10 \\
Actual region   & 36.2 & 2 &  6 & 19 & 16 & 33 & 27 \\
\\
\tableline
\end{tabular}
\end{center}
\end{table}




\end{document}